\documentclass[10pt, twocolumn, letterpaper, journal]{IEEEtran}
\usepackage{amssymb,amsmath,tabularx,delarray,enumerate,subfigure,graphicx,array,cite,color}
\usepackage{tikz,mathpazo}
\usetikzlibrary{shapes.geometric,arrows}
\usepackage{chemarr,float,array,booktabs,multirow,threeparttable}
\usepackage[english]{babel}

\usepackage[utf8]{inputenc}

\begin{document}
\title{FMO Study of the Interaction Energy between Human Estrogen Receptor $\alpha$ and Selected Ligands}
\author{Ricardo Ugarte$^\ast$
\thanks{$^\ast$Instituto de Ciencias Quimicas, Facultad de Ciencias, Universidad Austral de Chile. Independencia 641, Valdivia, Chile (e-mail: rugarte@uach.cl).}}
\markboth{december 2022}{}
\maketitle

\begin{abstract}

Fragment molecular orbital (FMO) calculations were performed in aqueous media which allowed us to obtain the interaction energy between the human estrogen receptor $\alpha$ ligand-binding domain (ER) and the selected ligands (L): 17$\beta$-estradiol (E2), 17$\alpha$-estradiol (17$\alpha$-E2), estriol (E3), genistein (GNT), diethylstilbestrol (DES), bisphenol A (BPA), bisphenol AF (BPAF), hydroxychlor (HPTE) and methoxychlor (DMDT). These calculations were carried out on representative structures of L-ER complexes obtained from molecular dynamics simulations. The MP2/6-31G(d) L-ER FMO interaction energy in kcal/mol is as follows: \\

\begin{scriptsize}
\scriptsize {E3 (-100.1) $<$ GNT (-95.8) $<$ E2 (-88.5) $<$ BPA (-84.7) $<$ DES (-82.6) $<$ BPAF (-80.6) $<$ 17$\alpha$-E2 (-78.7) $<$ HPTE (-75.9) $<$ DMDT (-46.3)}
\end{scriptsize}\\
\\
The central hydrophobic core of the ligands interacts attractively with several apolar amino acid residues of ER. Glu 353 and His 524 interacts strongly with most ligands through a hydrogen bond with the hydroxyl group of the phenol A-ring and the terminal hydroxylated ring, respectively. Water molecules were found at the binding site of receptor. In our model systems we have demonstrated what is generally observed in ligand-receptor complexes: the steric and chemical complementarity of the groups on the ligand and binding site surfaces.

\end{abstract}


\section{Introduction}

Estrogen receptors (ERs) are a group of nuclear proteins activated by 17$\beta$-estradiol (E2). ERs have relatively large ligand binding domains (LBD) and are promiscuous in terms of binding a wide variety of non-steroidal compounds. This can be attributed to the size of the ligand binding pocket, which has a accessible volume of 450 \r{A}$^3$ nearly twice that of the molecular volume of E2 (245 \r{A}$^3$). Since biological responses in target tissues are elicited by the interaction between substrate and cytoplasmic ER, a knowledge of the nature of this interaction is essential in any study of substrate response.

\begin{figure} [!htp]
\begin{center}
\includegraphics[width=1.0\columnwidth]{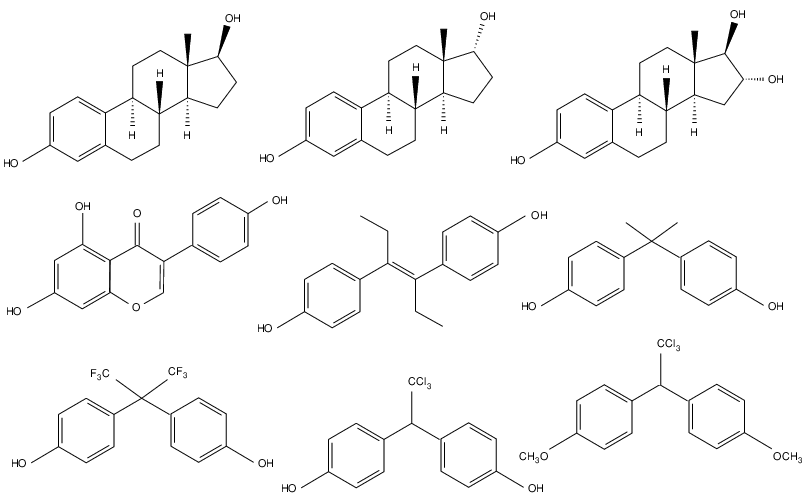}
\end{center}
        \setlength{\abovecaptionskip}{-3pt}
    \caption{Chemical structure of ligands. Top: 17$\beta$-Estradiol (left), 17$\alpha$-Estradiol (center), estriol (right). Middle: genistein (left), diethylstilbestrol (center), bisphenol A (right). Bottom: bisphenol AF (left), hydroxychlor (center), methoxychlor (right).}
    \label{Figure 1}
\end{figure}

Endocrine-disrupting chemicals (EDCs) can mimic estrogen action by bind to the receptor ligand binding site \cite{1,2}. From this point of view, the ER represents a good model for identifying and assessing the health risk of potential EDCs. This property characterizing EDCs could be reflected in the ligand-ER interaction energy. EDCs include synthetic chemicals used as: industrial solvents/lubricants and their byproducts (polychlorinated biphenyls, polybrominated biphenyls, dioxins), plastics (bisphenol A), plasticizers (phthalates), pesticides (methoxychlor, dichlorodiphenyltrichloroethane or DDT), fungicides (vinclozolin) and pharmaceutical agents (diethylstilbestrol). Natural chemicals found in human and animal food chains, e.g. phytoestrogens, including genistein and coumestrol, can also act as endocrine disruptors \cite{3}.

Several experimental and theoretical studies have been performed to investigate the ligand-ER interaction \cite{4,5,6,7,8,9,10,11,12,13,14,15,16,17,18,19,20,21,22,23,24,25,26,27,28,29,30,31}, and since 1997 about 360 crystal structures of ER LBD with different ligands have been solved and deposited in the Research Collaboratory for Structural Bioinformatics (RCSB) Protein Data Bank (PDB). On the basis of the above information, the mode of binding between ERs and their ligands has been determined. The specific recognition between ER and its ligand mainly depends on hydrogen bonds and hydrophobic interactions \cite{2,32,33}. Most of the theoretical studies which use the structures deposited in RCSB PDB have been carried out by means of molecular dynamics simulations (MD). Comparatively few studies have used quantum mechanical methods because they can be very computationally expensive and time consuming. However, there are hybrid methods that combine the precision of quantum mechanics and the speed of MD empirical force fields (QM/MM) \cite{34,35}. Another efficient alternative is the Fragment Molecular Orbital (FMO) method \cite{36,37,38}, which has been used for efficient and accurate QM calculations in very large molecular systems \cite{14,39}. FMO involve fragmentation of the chemical system, and from ab initio or density functional quantum mechanical calculations of each fragments (monomers) and their dimers (and trimers if greater accuracy is required) one can construct the total properties. The method includes the field of the full system in each individual fragment calculation and uses the systematic many-body expansion. The FMO method is suited to various analyses, as it provides information on fragments and their interactions that are naturally built into the method.

The present study aimed to investigate the relative binding affinities between the human estrogen receptor $\alpha$ ligand-binding domain (hER$\alpha$ LBD) and the following ligands (L) in aqueous medium (Figure 1):  \\
 i) Endogenous estrogens: 17$\beta$-estradiol (E2), 17$\alpha$-estradiol (17$\alpha$-E2), estriol (E3).   \\
ii) Phytoestrogens: genistein (GNT). \\
ii) Endocrine-disrupting chemicals: diethylstilbestrol (DES), bisphenol A (BPA), bisphenol AF (BPAF), hydroxychlor (HPTE), methoxychlor (DMDT).

\section{Methods}

The methods have been fully described elsewhere \cite{40} and so only a few of their salient features will be given here. Crystal structure of the HER$\alpha$ LBD in complex with E2 (PDB code 1G50) at 2.9 \r{A} resolution was retrieved from the RSCB PDB \cite{41} (Figure 2).

\begin{figure} [!htp]
\begin{center}
\includegraphics[width=1.0\columnwidth]{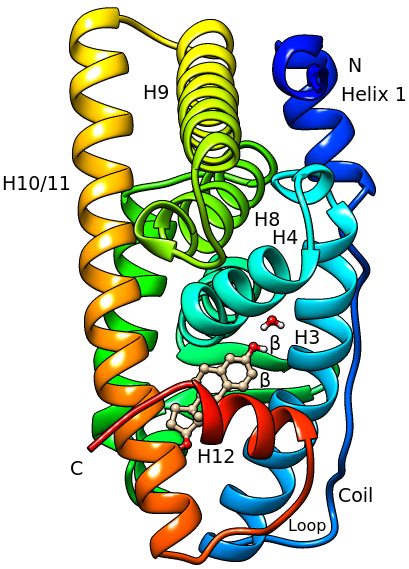}
\end{center}
        \setlength{\abovecaptionskip}{-3pt}
    \caption{Model of HER$\alpha$ LBD (ribbon). E2 and water molecule (ball and stick) at the binding site \cite{44}. The model based on the RSCB PDB crystal structure (PDB code 1G50) includes 247 amino acid residues.}
    \label{Figure 2}
\end{figure}

A model is built from 1G50 containing 247 amino acid residues and E2; 13394 TIP3P water molecules are included in the above model, resulting in E2-ER-W system. The rest of the systems are obtained from E2-ER-W, substituting E2 by the respective ligand (docking). 

The solvated systems (L-ER-W) were energy minimized according to a protocol that involved stages of steep descent and conjugate gradient. All MD simulations of the systems were carried out with the SANDER module of the AmberTools 15 \cite{42} with periodic boundary conditions, using Particle Mesh Ewald method to treat long-range electrostatics interactions with a non-bonded cutoff of 10 \r{A}. All bonds involving hydrogen atoms were restrained using the SHAKE algorithm. Temperature regulation was done using a Langevin thermostat with collision frequency of 1 $ps^{-1}$. The Berendsen barostat was used for constant pressure simulation at 1 atm, with a relaxation time of 1 ps. The time step was 1 or 2 fs. The hydrogen mass distribution (HMR) method was used for accelerating MD simulations [43]. The energy-minimized systems was submitted to the following protocol:
\\

\tikzstyle{startstop} = [rectangle, rounded corners, minimum width=3cm, minimum height=1cm,text centered, draw=black, fill=red!30]
\tikzstyle{arrow} = [->,>=stealth]

Scheme 1:
\\
\begin{center}
\begin{tikzpicture}[node distance=2cm]
\node (start) [startstop] {NVT: 0 $\rightarrow$ 310 K \ $ \Delta t = 1 $ fs \ 100 ps};
\node (in1) [startstop, below of=start] {NPT: 310 K \ $ \Delta t = 2 $ fs \ 500 ps};
\node (pro1) [startstop, below of=in1] {NVT: 310 $\rightarrow$ 5 K \ $ \Delta t = 1 $ fs \ 100 ps};
\draw [arrow](start) -- (in1);
\draw [arrow](in1) -- (pro1);
\end{tikzpicture}
\end{center}
$  $
\\
From the restart file of the last simulation in Scheme 1, we performed an extensive set of molecular dynamics simulations to explore the conformational space in the vicinity of the crystallographic structure. To circumvent the limited conformational sampling ability of MD simulations at 310 K, we used multiple-trajectory short-time simulations \cite{45}. By combining the sampling ability of the multiple trajectories, we expect to sample more conformational space than single trajectory of the same length. The aforementioned restart file was used as seed for 30 short-time simulations that obey the protocol established in Scheme 2:
\\

Scheme 2:
\\
\begin{center}
\begin{tikzpicture}[node distance=2cm]
\node (start) [startstop] {NVT: 5 $\rightarrow$ 150 K \ $ \Delta t = 2 $ fs \ 60 ps};
\node (in1) [startstop, below of=start] {NPT: 150 $\rightarrow$ 310 K \ $ \Delta t = 2 $ fs \ 140 ps};
\node (pro1) [startstop, below of=in1] {NVE: 310 K \ $ \Delta t = 2 $ fs \ 500 ps};
\draw [arrow](start) -- (in1);
\draw [arrow](in1) -- (pro1);
\end{tikzpicture}
\end{center}
$  $
\\
The initial velocities (Scheme 2) were assigned randomly from a Maxwell-Boltzmann distribution at 5 K. The trajectories start with the same structure and differ only in the initial velocity assignment. At the end of the equilibration, from $\sim$ 80 ps NPT ensemble, the average temperature of the final 60 ps was 310 K, and the average density was 1.0 g/mL. All production runs of 0.5 ns were performed in an NVE ensemble at 310 K.

A clustering approach for each system, based on the C$\alpha$-RMSD (root mean square deviation) was applied to the snapshots of the MD simulations \cite{46,47}. Prior to clustering, the individual trajectories from the 30 short-time simulations were combined into a single file (full trajectory), and the water molecules were removed to speed up the calculations. In our analysis, 7500 snapshots were grouped into three or five clusters. Thus, three or five representative structures (RSs) of each cluster, and therefore of the conformation population, were obtained (Table I).

All L-ER-W systems representative of the population were subjected to geometry optimization using Gaussian
09 at the MM/AMBER level of theory \cite{48}. In the next step, the water molecules beyond 10 \r{A} of the protein surface were deleted using VMD program \cite{49}. Thus, new representative models (L-ER-w) with a water layer of 10 \r{A} around of receptor surface were generated. ONIOM, \cite{34} a hybrid QM/MM method implemented in Gaussian 09, was used for the geometry optimization of L-ER-w models. In the present study we used a two-layer ONIOM(B3LYP/6-31G(d):AMBER) scheme: L(B3LYP/6-31G(d)); ER-w(AMBER).

The L-ER-w optimized structures (RS) were subjected to FMO calculations. The ligand-receptor FMO interaction energy (E$_{int}$) is based on obtaining and sum the interaction energies between all pairs of fragments ligand-amino acid residue and if applicable, ligand-water trapped on binding site. The AFO (adaptive frozen orbitals) scheme was used throughout for fragmentation across peptide bonds, with the default settings for bond definitions. The fragmentation of the model was as follows: the first two amino acid residues and each remaining amino acid residue of apo-ER, L and the water molecule were treated as a single fragment. The RS was divided into two layers treated at different levels of theory: FMO2-RHF/STO-3G:MP2/6-31G(d). The layer 1 (aqueous environment) described by RHF/STO-3G and layer 2 (L-ER) described by MP2/6-31G(d) \cite{50}. The water molecules of the binding site were included in the layer 2. Finally, pair interaction energies were computed in all RSs.

\section{Results and Discussion}

\subsection{Cluster Analysis of MD trajectories}

Table I show the clustering results of MD trajectories. In general, at the specified simulation time (0.5 ns), the clustering results showed that 1-3 clusters are sufficient to represent at least 88\% of the population of conformations in all systems. The only exception is E2-ER-W system which requires four clusters. The ER-W (apo-receptor) required three clusters and the remaining systems one or two. Coincidence, or is it a feature of the natural substrate to display an equal or greater diversity of conformations than the apo-receptor?
  
\subsection{The Binding Site}

The definition of the binding site is somewhat arbitrary. For our purposes, we define the binding site as those amino acid residues in direct contact with the ligand, i.e. all receptor residues that are within a distance of 4.0 \r{A} of the ligand (Figures 3-6). According to the simulation results, the binding site is a "dynamic entity" that depends on the ligand under consideration. Tables II-III support the above statement. The canonical binding site is defined as the binding site associated with the natural substrate 17$\beta$-estradiol (E2). In the following discussion, we will assume that each RS reliably describes the main features of each and every conformation of the cluster to which it belongs, in particular, the binding site. Based on the above affirmation, the canonical amino acid residues present in the RSs of each cluster are shown in boldface type (Tables II-III). Phe 425 is included in the canonical binding site as it is "observed" in $\approx$ 60\% of the population of E$_{2}$-ER conformations.

\begin{table*}[!htbp]
 \begin{center}
   \resizebox{\textwidth}{!}{%
  \begin{threeparttable}
\caption{Cluster Analysis of MD trajectories in the different systems}
\begin{tabular}{cccccccccccccccc}
  \toprule[1pt]
  & \multicolumn{5}{c}{{E2-ER-W}}
  & \multicolumn{7}{c}{{17$\alpha$-E2-ER-W}}
  & \multicolumn{3}{c}{{E3-ER-W}} \\
  \midrule[1pt]
    Cluster & 1 & 2 & 3 & 4 & 5 & & 1 & 2 & 3 & 4 & 5 & & 1 & 2 & 3 \\
  CP$^{(a)}$ & 3630 & 1749 & 1169 & 907 & 45 & & 5156 & 185 & 543 & 77 & 1539 & \ \ \ & 7253 & 145 & 102 \\
  CP \% & 48.4 & 23.3 & 15.6 & 12.1 & 0.6 & & 68.8 & 2.5 & 7.2 & 1.0 & 20.5 & \ \ \ & 96.7 & 1.9 & 1.4 \\
  RS$^{(b)}$ & \ $E2_{1}$ & $E2_{2}$ & $E2_{3}$ & $E2_{4}$ & $E2_{5}$ & & 17$\alpha E2_{1}$ & 17$\alpha E2_{2}$ & 17$\alpha E2_{3}$ & 17$\alpha E2_{4}$ & 17$\alpha E2_{5}$ & \ \ \  & $E3_{1}$ & $E3_{2}$ & $E3_{3}$ \\
  \toprule[1pt]
  & \multicolumn{5}{c}{{GNT-ER-W}}
  & \multicolumn{7}{c}{{DES-ER-W}}
  & \multicolumn{3}{c}{{BPA-ER-W}} \\
  \midrule[1pt]
    Cluster & & 1 & 2 & 3 & & & & 1 & 2 & 3 & & \ \ \ & 1 & 2 & 3 \\
  CP \ \ \ & & 6392 & 919 & 189 & & & & 6608 & 601 & 291 & & \ \ \ & 6018 & 1423 & 59 \\
  CP \% & & 85.2 & 12.3 & 2.5 & & & & 88.1 & 8.0 & 3.9 & & \ \ \ & 80.2 & 19.0 & 0.8 \\
  RS \ \ \ & \ & $GNT_{1}$ & $GNT_{2}$ & $GNT_{3}$ & & & & $DES_{1}$ & $DES_{2}$ & $DES_{3}$ & & \ \ \ & $BPA_{1}$ & $BPA_{2}$ & $BPA_{3}$ \\
  \toprule[1pt]
  & \multicolumn{5}{c}{{BPAF-ER-W}}
  & \multicolumn{7}{c}{{HPTE-ER-W}}
  & \multicolumn{3}{c}{{DMDT-ER-W}} \\
  \midrule[1pt]
    Cluster & & 1 & 2 & 3 & & & & 1 & 2 & 3 & & \ \ \ & 1 & 2 & 3 \\
  CP \ \ \ & & 7212 & 144 & 144 & & & & 6500 & 878 & 122 & & \ \ \ & 5928 & 1414 & 158 \\
  CP \% & & 96.2 & 1.9 & 1.9 & & & & 86.7 & 11.7 & 1.6 & & \ \ \ & 79.0 & 18.9 & 2.1 \\
  RS \ \ \ & \ & $BPAF_{1}$ & $BPAF_{2}$ & $BPAF_{3}$ & & & & $HPTE_{1}$ & $HPTE_{2}$ & $HPTE_{3}$ & & \ \ \ & $DMDT_{1}$ & $DMDT_{2}$ & $DMDT_{3}$ \\
  \bottomrule[1pt]
\end{tabular}
\label{Table I}
\begin{tablenotes}
      \footnotesize
      \item $^{(a)}$ Cluster population. $^{(b)}$ Representative structure of the respective cluster.
\end{tablenotes}
  \end{threeparttable}%
   }
    \end{center}
\end{table*}

\begin{table*}[!htbp]
 \begin{center}
  \resizebox{\textwidth}{!}{%
   \begin{threeparttable}
    \caption{Ligand-Amino Acid FMO Interaction Energy}
\begin{tabular}{ccccccccccccccccc}
  \toprule[1pt]  
  & \multicolumn{1}{c}{{$E2_{1}$}}
  & \multicolumn{1}{c}{{$E2_{2}$}}
  & \multicolumn{1}{c}{{$E2_{3}$}}
  & \multicolumn{1}{c}{{$E2_{4}$}}
  & \multicolumn{1}{c}{{$E2_{5}$}}
  & \multicolumn{1}{c}{{17$\alpha E2_{1}$}}
  & \multicolumn{1}{c}{{17$\alpha E2_{2}$}}
  & \multicolumn{1}{c}{{17$\alpha E2_{3}$}}
  & \multicolumn{1}{c}{{17$\alpha E2_{4}$}}
  & \multicolumn{1}{c}{{17$\alpha E2_{5}$}}
  & \multicolumn{1}{c}{{$E3_{1}$}}
  & \multicolumn{1}{c}{{$E3_{2}$}}
  & \multicolumn{1}{c}{{$E3_{3}$}}
  & \multicolumn{1}{c}{{$GNT_{1}$}}
  & \multicolumn{1}{c}{{$GNT_{2}$}}
  & \multicolumn{1}{c}{{$GNT_{3}$}} \\
  \midrule[1pt]
  \textbf{Met 343} & -3.32 & -4.26 & -2.52 & -3.42 & -2.30 & -0.99 & -3.51 & -1.68 & -1.60 & -1.90 & -2.11 & -2.44 & -2.61 & -3.13 & -2.48 & -2.18 \\
  \textbf{Leu 346} & -2.55 & -2.40 & -2.30 & -2.56 & -2.18 & -1.94 & -1.54 & -1.98 & -2.05 & -2.08 & -1.66 & -1.98 & -2.05 & -2.45 & -2.62 & -1.60 \\
  \textbf{Thr 347} & -4.43 & -3.83 & -4.97 & -4.25 & -4.37 & -1.16 & -3.74 & -5.97 & -5.31 & -3.02 & -2.06 & -5.19 & -5.78 & -8.98 & -10.64& -9.23 \\
  \textbf{Leu 349} & -0.82 & -1.10 & -0.09 & -0.99 & -0.57 & -2.66 & -2.83 & -0.91 & -0.30 & -2.79 & -2.76 & -0.64 & -0.86 &  0.02 & -0.38 & -0.43 \\
  \textbf{Ala 350} & -2.43 & -2.27 & -2.26 & -2.11 & -2.55 & -1.90 & -0.94 & -2.62 & -4.28 & -2.37 & -2.31 & -2.28 & -2.60 & -1.57 & -1.75 & -1.51 \\
  \textbf{Glu 353} & -24.51& -24.49& -24.21& -23.16& -22.92& -22.55& -27.37& -21.00& -2.24 & -24.92& -24.91& -22.76& -16.99& -26.61& -25.78& -26.56 \\
  Trp 383          &       &       &       &       &       &       &       & -0.21 &       &       &       &       &       &       &       &       \\
  \textbf{Leu 384} & -1.41 & -1.66 & -1.65 & -1.33 & -1.02 & -2.15 & -2.77 & -1.85 & -2.01 & -2.66 & -2.72 & -1.78 & -1.97 & -1.36 & -1.65 & -1.52 \\
  \textbf{Leu 387} & -3.69 & -3.49 & -2.50 & -3.99 & -2.88 & -2.12 & -2.84 & -5.10 & -2.66 & -2.39 & -2.30 & -3.53 & -1.96 & -3.13 & -3.54 & -2.90 \\
  \textbf{Met 388} & -0.54 & -1.00 & -0.50 &  0.02 & -2.14 & -5.75 & -5.06 & -0.46 & -1.84 & -5.20 & -4.98 & -0.92 & -1.44 & -1.46 & -0.44 & -1.19 \\
  \textbf{Leu 391} & -5.41 & -4.68 & -4.86 & -4.67 & -3.71 & -2.93 & -2.64 & -3.16 & -3.88 & -3.06 & -3.30 & -3.89 & -4.00 & -5.35 & -4.55 & -5.16 \\
  \textbf{Arg 394} & 3.24  & 4.40  &  2.45 &  2.56 &  2.55 &  1.00 &  1.04 &  1.76 & -9.07 & -0.06 &  0.07 & 0.94 &  0.68 &  4.88 &  4.18 &  4.64 \\
  Leu 402          &       &       &       &       &       &       &       &       &       &       &       &       &       &       &       &       \\
  \textbf{Phe 404} & -4.55 & -4.62 & -4.15 & -3.96 & -3.79 & -4.72 & -3.67 & -4.09 & -4.02 & -5.03 & -4.95 & -3.82 & -3.19 & -4.20 & -3.06 & -3.07 \\
  Val 418          &       & -0.15 &       &       &       &       &       &       &       &       &       &       & -0.33 &       &       &       \\
  \textbf{Met 421} & -2.75 & -2.23 & -2.62 & -3.15 & -2.48 & -4.29 & -3.88 & -2.41 & -3.17 & -3.55 & -4.50 & -5.05 & -5.39 & -2.56 & -2.70 & -1.92 \\
  \textbf{Ile 424} & -1.43 & -1.57 & -1.27 & -1.36 & -1.02 & -2.91 & -1.26 & -1.05 & -1.31 & -1.27 & -3.42 & -3.29 & -2.99 & -1.71 & -2.69 & -3.25 \\
  \textbf{Phe 425} & -1.31 &   x    &    x   &  0.07 & -0.06 & -0.15 &   x    &   x    &   x    &   x    &       x &   x    &   x    &   x    &   x    &   x    \\
  \textbf{Leu 428} & -1.11 & -1.13 & -1.09 & -1.25 & -0.88 & -1.13 & -1.39 & -1.15 & -0.71 & -0.89 & -1.02 & -1.40 & -1.35 &   x   &   x    &    x   \\
  \textbf{Gly 521} & -1.23 & -0.86 & -1.07 & -1.21 & -0.80 & -0.80 & -0.50 & -0.38 & -0.04 & -0.43 & -1.70 & -1.46 & -1.68 & -1.80 & -0.91 & -1.39 \\
  \textbf{His 524} & -16.82& -16.42& -16.45& -15.60& -16.61& -15.86& -13.09& -11.96& -14.01& -13.41& -26.47& -26.60& -27.11& -23.29& -23.60& -24.02 \\
  \textbf{Leu 525} & -4.11 & -4.03 & -3.26 & -3.97 & -2.73 & -1.84 & -1.93 & -2.56 & -2.45 & -1.72 & -4.16 & -4.41 & -4.53 & -6.72 & -6.05 & -5.48 \\
  \textbf{Met 528} & 0.33  & 0.06  & -0.54 &  0.19 & -1.00 &   x    &   x    &  0.62 &   x    &   x    &       x &  0.24 &  0.34 & -0.30 & -0.22 & -0.78 \\
  Val 534          &       &       &       &       &       &       &       &       &       &       &       &       &       & -0.39 & -0.70 & -0.32 \\
  Leu 540          &       &       &       &       & -0.38 &       &       & -0.41 & -0.38 &       & -0.41 &       &       &       &       &       \\
  \midrule[1pt]  
  $E_{L-Aa(BS)}$    & -78.85 &-75.73& -73.86& -74.14 & -71.84& -74.85& -77.92& -66.57& -61.33& -76.75&  -95.67& -90.26& -85.86& -90.11& -89.58 & -87.87 \\      
  \bottomrule[1pt]
\end{tabular}
\label{Table II}
\begin{tablenotes}
      \footnotesize
      \item Boldface type indicates canonical amino acid residues. All energies at MP2/6-31G(d) level of theory in kcal/mol.
\end{tablenotes}
  \end{threeparttable}%
   }
    \end{center}
\end{table*}

The RSs of each selected ligands contain binding sites that display changes in amino acid residue composition with respect to the canonical binding site. If we analyze those changes (Table IV) we can infer the existence of missing and extra residues. The following residues is found to be missing: Met 528 (65.5\%), Phe 425 (39.2\%) and Leu 428 (22.2\%) of the binding site of RSs. The remaining missing residues account for less than 1.6\%. Regarding extra residues the occurrences are as follows: Leu 540 (47.8\%), Trp 383 (36.4\%), Val 534 (22.2\%), Leu 402 (20.9\%) and Val 418 (2.7\%). In general, the interaction energy between these residues and the ligand is very low.

Diethylstilbestrol (DES), is a very potent synthetic non-steroidal selective estrogen receptor modulator \cite{51}. DES RSs are the only that preserve the canonical binding site residues. Maybe, this feature is critical for the high affinity to the receptor. DES RSs integrates three extra residues to the binding site (Trp 383, Val 534 and Leu 540).

Genistein (GNT), is a phytoestrogen with lower potency than E2 \cite{52}. GNT RSs retains Met 528 at the binding site. Although GNT resembles DES ligand, it does not retain Phe 425 and Leu 428. GNT RSs integrates Val 534 extra residues to the binding site.

Selected steroidal estrogens show affinity for the estrogen receptor. The order of potency is as follows: E2 $>$ E3 $>$ 17$\alpha$-E2 \cite{53}. Met 528 is missing at the binding site of the two main 17$\alpha$E2 RSs and $E3_{1}$. Phe 425 is missing at the binding site of E3 RSs, 17$\alpha$-E2$_{2-5}$ and E2$_{2-3}$. Leu 428 is conserved at the binding site of selected steroidal estrogens RSs. 17$\alpha$-E2$_{3}$ integrates Trp 383 and Leu 540. $E2_{2}$ and $E3_{1}$ integrates, respectively, Val 418 and Leu 540 extra residues to the binding site.

\begin{table*}[!htbp]
 \begin{center}
  \resizebox{\textwidth}{!}{%
   \begin{threeparttable}
    \caption{Ligand-Amino Acid FMO Interaction Energy (Cont.)}
\begin{tabular}{cccccccccccccccc}
  \toprule[1pt]  
  & \multicolumn{1}{c}{{$DES_{1}$}}
  & \multicolumn{1}{c}{{$DES_{2}$}}
  & \multicolumn{1}{c}{{$DES_{3}$}}
  & \multicolumn{1}{c}{{$BPA_{1}$}}
  & \multicolumn{1}{c}{{$BPA_{2}$}}
  & \multicolumn{1}{c}{{$BPA_{3}$}}
  & \multicolumn{1}{c}{{$BPAF_{1}$}}
  & \multicolumn{1}{c}{{$BPAF_{2}$}}
  & \multicolumn{1}{c}{{$BPAF_{3}$}}
  & \multicolumn{1}{c}{{$HPTE_{1}$}}
  & \multicolumn{1}{c}{{$HPTE_{2}$}}
  & \multicolumn{1}{c}{{$HPTE_{3}$}}
  & \multicolumn{1}{c}{{$DMDT_{1}$}}
  & \multicolumn{1}{c}{{$DMDT_{2}$}}
  & \multicolumn{1}{c}{{$DMDT_{3}$}} \\
  \midrule[1pt]
  \textbf{Met 343} & -2.45 & -2.40 & -2.38 & -0.59 & -3.66 & -0.36 & -2.00 & -1.32 & -3.38 & -0.52 & -1.78 & -2.58 & -1.46 & -1.02 & -1.42 \\
  \textbf{Leu 346} & -1.98 & -2.41 & -2.99 & -1.33 & -2.16 & -2.03 & -2.83 & -2.70 & -2.89 & -3.91 & -3.08 & -2.95 & -3.64 & -2.20 & -1.71 \\
  \textbf{Thr 347} & -4.64 & -4.62 & -4.24 & -4.88 & -4.16 & -5.92 & -7.42 & -5.04 & -4.66 & -1.41 & -1.80 & -0.82 & -5.77 & -1.74 & -2.34 \\
  \textbf{Leu 349} & -0.60 & -0.92 & -0.18 & -0.57 & -0.86 & -0.03 &  0.19 & -0.97 &   x    & -2.62 & -2.87 & -2.69 & -0.61 & -2.39 &   x    \\
  \textbf{Ala 350} & -2.24 & -2.35 & -2.20 & -1.88 & -1.23 & -2.02 & -3.80 & -2.38 & -2.85 & -1.90 & -2.28 & -3.22 & -2.68 & -3.50 & -2.67 \\
  \textbf{Glu 353} & -22.40& -23.57& -24.77& -26.89& -3.54 & -28.13& -5.34 & -5.91 &   x    & -30.54& -31.59& -31.25& -5.03 & -7.31 & -8.11 \\
  Trp 383          & -0.58 & -0.84 & -0.79 & -0.63 & -0.31 & -0.68 &       &       & -0.35 & -0.87 & -1.34 & -1.26 &       & -1.01 &       \\
  \textbf{Leu 384} & -0.44 & -0.86 & -1.22 & -1.33 & -1.63 & -1.43 & -2.63 & -1.27 & -1.78 & -1.08 & -1.37 & -0.96 & -1.64 & -1.44 & -1.58 \\
  \textbf{Leu 387} & -4.14 & -3.66 & -3.96 & -4.80 & -2.57 & -4.47 & -3.21 & -3.38 & -2.90 & -2.86 & -2.71 & -2.37 & -3.23 & -1.72 & -3.89 \\
  \textbf{Met 388} &  0.56 &  0.58 &  0.39 & -0.33 & -6.75 & -0.22 & -3.88 & -10.09& -5.90 & -8.98 & -9.52 & -6.18 & -3.52 & -6.96 & -3.52 \\
  \textbf{Leu 391} & -4.76 & -4.56 & -4.90 & -4.63 & -2.99 & -4.07 & -3.37 & -2.17 & -2.38 & -3.06 & -3.49 & -3.10 & -2.75 & -1.94 & -4.56 \\
  \textbf{Arg 394} &  2.38 &  3.60 &  3.42 &  3.15 & -0.20 &  1.76 &  1.45 &  1.83 &   x    &  5.73 &  5.43 &  4.95 &  2.41 &  3.26 &  4.00 \\
  Leu 402          &       &       &       &       &       &       & -0.31 &       &       &       & -0.49 & -0.40 & -0.58 &       &       \\
  \textbf{Phe 404} & -3.80 & -3.84 & -4.20 & -4.07 & -4.41 & -3.97 & -2.12 & -3.98 & -2.83 & -4.44 & -6.98 & -7.55 & -2.27 & -2.92 & -3.15 \\
  Val 418          &       &       &       &       &       &       &       &       &       &       &       &       &       &       &       \\
  \textbf{Met 421} & -1.29 & -1.06 & -2.28 & -4.66 & -2.92 & -3.96 & -19.69 & -3.45 & -4.26 & -1.32 & -6.48 & -7.84 & -6.01 & -4.04 & -5.14 \\
  \textbf{Ile 424} & -1.27 & -1.06 & -0.94 & -2.34 & -2.58 & -2.43 & -2.02 & -3.32 & -3.10 & -3.07 & -1.00 & -1.45 & -2.76 & -3.41 & -3.00 \\
  \textbf{Phe 425} & -0.12 &  0.01 & -0.19 &   x    & -0.44 &    x   &  1.69 &   x    &   x   & -0.01 & -1.94 & -1.07 & -1.36 & -2.58 & -1.56 \\
  \textbf{Leu 428} & -0.46 & -0.60 & -0.67 &   x    &    x   &   x    & -1.90 & -0.70 & -0.85 & -1.48 & -1.25 & -0.93 & -1.12 & -0.14 & -1.13 \\
  \textbf{Gly 521} & -1.65 & -1.43 & -1.32 & -1.25 & -1.33 & -1.18 & -1.79 & -1.65 & -1.41 & -1.01 &       x & -0.20 & -0.54 & -0.37 & -0.24 \\
  \textbf{His 524} & -16.00& -17.66& -19.50& -16.08& -15.62& -15.17& -6.26 & -14.20& -16.15& -6.58 &       x & -0.04 & -1.56 & -1.52 &   x    \\
  \textbf{Leu 525} & -5.61 & -7.00 & -4.70 & -2.84 & -3.14 & -2.72 & -2.42 & -3.09 & -4.58 & -1.66 & -1.89 & -1.88 & -0.80 & -1.34 & -1.12 \\
  \textbf{Met 528} & -2.01 & -0.95 & -1.07 &   x    &   x    &   x    &   x    &   x    &   x    &   x    &   x    &   x    &   x    &   x    &   x    \\
  Val 534          & -0.80 & -1.10 & -0.79 &       &       &       &       &       &       &       &       &       &       &       &       \\
  Leu 540          & -0.70 & -0.88 & -0.88 & -0.85 & -0.67 & -1.00 &       & -1.87 & -0.46 & -0.54 & -0.73 & -0.76 &       & -0.91 & -0.52 \\
  \midrule[1pt]  
  $E_{L-Aa(BS)}$    & -75.00 &-77.58& -80.36& -76.80 & -61.17& -78.03& -67.66& -65.66& -60.73& -72.13&  -77.16& -74.55& -44.92& -45.20& -41.66 \\      
  \bottomrule[1pt]
\end{tabular}
\label{Table III}
\begin{tablenotes}
      \footnotesize
      \item Boldface type indicates canonical amino acid residues. All energies at MP2/6-31G(d) level of theory in kcal/mol.
\end{tablenotes}
  \end{threeparttable}%
   }
    \end{center}
\end{table*}

\begin{table*}[!htbp]
 \begin{center}
   \resizebox{\textwidth}{!}{%
  \begin{threeparttable}
\caption{Representative Structures containing Modified Amino Acid Residues at the Binding Site$^{(a)}$}
\begin{tabular}{lccccccccccccccccccc}
  \toprule[1pt]
  & \multicolumn{1}{c}{{}}
  & \multicolumn{14}{c}{{Representative Structure}} \\
Missing Residues & & & & & & & & & & & & & & & & & & \\
  \midrule[1pt]
\ \ \ \ \ Leu 349 & $BPAF_{3}$ & $DMDT_{3}$ & & & & & & & & & & & & & & & & \\
\ \ \ \ \ Glu 353 & $BPAF_{3}$ & & & & & & & & & & & & & & & \\
\ \ \ \ \ Arg 394 & $BPAF_{3}$ & & & & & & & & & & & & & & & \\
\ \ \ \ \ Phe 425 & $E2_{2}$ & $E2_{3}$ & 17$\alpha E2_{2}$ & 17$\alpha E2_{3}$ & 17$\alpha E2_{4}$ & 17$\alpha E2_{5}$ & $E3_{1}$ & $E3_{2}$ & $E3_{3}$ & $GNT_{1}$ & $GNT_{2}$ & $GNT_{3}$ & $BPA_{1}$ & $BPA_{3}$ & $BPAF_{2}$ & $BPAF_{3}$ & & \\
\ \ \ \ \ Leu 428 & $GNT_{1}$ & $GNT_{2}$ & $GNT_{3}$ & $BPA_{1}$ & $BPA_{2}$ & $BPA_{3}$ & & & & & & & & & & & & \\
\ \ \ \ \ Gly 521 & $HPTE_{2}$ & & & & & & & & & & & & & & & & & \\
\ \ \ \ \ His 524 & $HPTE_{2}$ & $DMDT_{3}$ & & & & & & & & & & & & & & & & \\
\ \ \ \ \ Met 528 & 17$\alpha E2_{1}$ & 17$\alpha E2_{2}$ & 17$\alpha E2_{4}$ & 17$\alpha E2_{5}$ & $E3_{1}$ & $BPA_{1}$ & $BPA_{2}$ & $BPA_{3}$ & $BPAF_{1}$ & $BPAF_{2}$ & $BPAF_{3}$ & $HPTE_{1}$ & $HPTE_{2}$ & $HPTE_{3}$ & $DMDT_{1}$ & $DMDT_{2}$ & $DMDT_{3}$ & \\
  \toprule[1pt]
  & \multicolumn{1}{c}{{}}
  & \multicolumn{14}{c}{{Representative Structure}} \\
Extra Residues & & & & & & & & & & & & & & & & & & \\
  \midrule[1pt]
\ \ \ \ \ Trp 383 & 17$\alpha E2_{3}$ & $DES_{1}$ & $DES_{2}$ & $DES_{3}$ & $BPA_{1}$ & $BPA_{2}$ & $BPA_{3}$ & $BPAF_{3}$ & $HPTE_{1}$ & $HPTE_{2}$ & $HPTE_{3}$ & $DMDT_{2}$ & & & & & & \\
\ \ \ \ \ Leu 402 & $BPAF_{1}$ & $HPTE_{2}$ & $HPTE_{3}$ & $DMDT_{1}$ & & & & & & & & & & & & & & \\
\ \ \ \ \ Val 418 & $E2_{2}$ & $E3_{3}$ & & & & & & & & & & & & & & & & \\
\ \ \ \ \ Val 534 & $GNT_{1}$ & $GNT_{2}$ & $GNT_{3}$ & $DES_{1}$ & $DES_{2}$ & $DES_{3}$ & & & & & & & & & & & & \\
\ \ \ \ \ Leu 540 & $E2_{5}$ & 17$\alpha E2_{3}$ & 17$\alpha E2_{4}$ & $E3_{1}$ & $DES_{1}$ & $DES_{2}$ & $DES_{3}$ & $BPA_{1}$ & $BPA_{2}$ & $BPA_{3}$ & $BPAF_{2}$ & $BPAF_{3}$ & $HPTE_{1}$ & $HPTE_{2}$ & $HPTE_{3}$ & $DMDT_{2}$ & $DMDT_{3}$ & \\
  \toprule[1pt]
  & \multicolumn{16}{c}{{Conserved Residues$^{(b)}$}} \\
& & & & & & & & & & & & & & & & & & & \\   
  \midrule[1pt] 
& Met 343 & Leu 346 & Thr 347 & Leu 349 & Ala 350 & Glu 353 & Leu 384 & Leu 387 & Met 388 & Leu 391 & Arg 394 & Phe 404 & Met 421 & Ile 424 & Leu 428$^{(c)}$ & Leu 525 & Gly 521 & His 524 & Leu 525\\
  \bottomrule[1pt]
\end{tabular} 
\label{Table IV}
\begin{tablenotes}
      \footnotesize
      \item $^{(a)}$ Reference: Canonical binding site. $^{(b)}$ Missing residues found in scarcely populated RS are considered as conserved residues. $^{(c)}$ Semi-conserved residue.
\end{tablenotes}
  \end{threeparttable}%
   }
    \end{center}
\end{table*}

The four selected EDCs \cite{54} bisphenol A (BPA), bisphenol AF (BPAF), hydroxychlor (HPTE), methoxychlor (DMDT) related to DDT, are chemically similar in terms of possessing two benzene rings with some functional groups at their para-positions. These has a lower potency than E2 \cite{55,56} and have a smaller structure than the above ligands, occupying less space at the binding site. Met 528 is missing at the binding site in the four EDCs RSs, Leu 428 is only missing in BPA RSs and Phe 425 is missing in $BPA_{1}$. All RSs do not contain Val 418 and Val 534 extra residues in their binding sites. $BPAF_{1}$ and $DMDT_{1}$ integrates Leu 402, BPA and HPTE RSs integrates Trp 383 and Leu 540 extra residues to the binding site.

The perturbation caused by the ligand at the binding site can be identified in terms of missing or extra residues. It is expected that less distortion of the canonical binding site will ensure better behavior of the ligand-receptor complex, since under this assumption the biological response of the natural substrate would be imitated.

In order to quantify the perturbation caused by missing and extra residues in a given system, a simple factor was derived: (i) each missing or extra residue of a RS is assigned a score of 1 or 3, respectively; (ii) these scores are multiplied by the respective cluster population (CP \%) and added together; (iii) this procedure is performed for all RSs constituting the system (5 or 3); (iv) The results obtained for each of the RSs are summed and the total is divided by the total obtained for the canonical structure, obtaining the distortion factor for the system. Based on these factors, the following order of perturbation of the binding site caused by the ligands is deduced: \\

\begin{scriptsize}
\scriptsize {E2 $<$ 17$\alpha$-E2 $<$ BPAF $<$ DMDT $<$ E3 $\simeq$ GNT $<$ BPA $<$ HPTE $<$ DES}
\end{scriptsize}\\
\\
with the synthetic non-steroidal DES being the most perturbing to the binding site and the steroidal estrogen 17$\alpha$-E2 the most similar in behavior to E2. In the derivation of the distortion factor it is assumed that an extra residue is more disruptive to the binding site than a missing one.

\begin{figure*} [!htp]
\begin{center}
\includegraphics[width=1.05\linewidth]{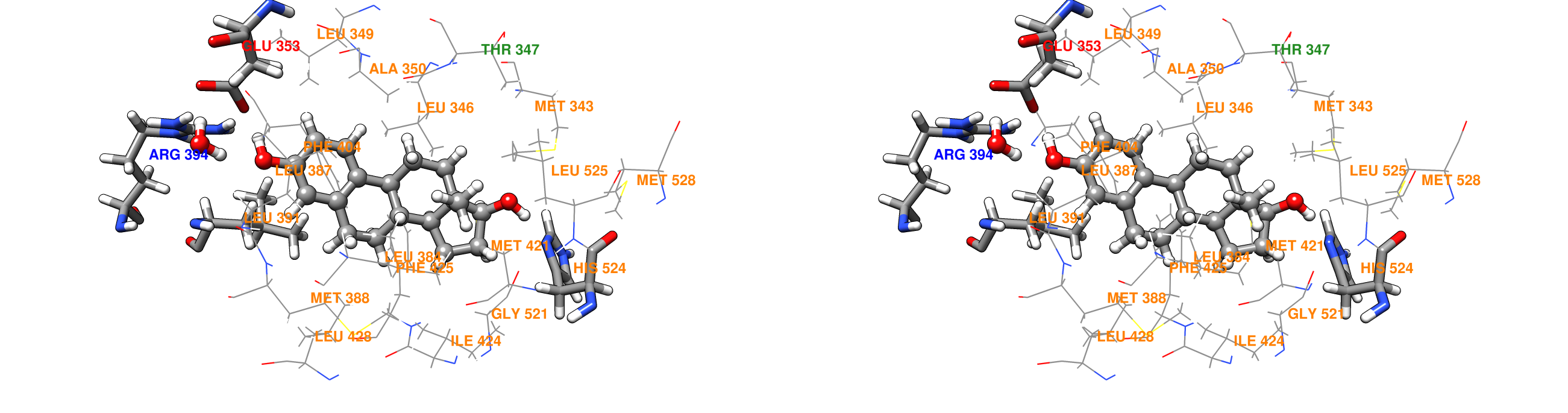}
\end{center}
        \setlength{\abovecaptionskip}{-16pt}
    \caption{Stereo view of canonical binding site: E2 (ball and stick) is shown surrounded by a hydrophobic pocket (wire), water molecule (ball and stick) and important amino acid residues (stick): Glu 353, Leu 391, Arg 394 and His 524. The image corresponds to $E2_{1}$ RS. Amino acid letter code: orange (apolar), green (polar), red (negatively charged), blue (positively charged).}
    \label{Figure 3}
\end{figure*}

\begin{table*}[!htbp]
 \begin{center}
   \resizebox{\textwidth}{!}{%
  \begin{threeparttable}
\caption{FMO Interaction Energy}
\begin{tabular}{lccccccccccccccc}
  \toprule[1pt]
  & \multicolumn{5}{c}{{E2-ER-w}}
  & \multicolumn{7}{c}{{17$\alpha$-E2-ER-w}}
  & \multicolumn{3}{c}{{E3-ER-w}} \\
  \midrule[1pt]
  \ \ \ RS & $E2_{1}$ & $E2_{2}$ & $E2_{3}$ & $E2_{4}$ & $E2_{5}$ & & 17$\alpha E2_{1}$ & 17$\alpha E2_{2}$ & 17$\alpha E2_{3}$ & 17$\alpha E2_{4}$ & 17$\alpha E2_{5}$ & & $E3_{1}$ & $E3_{2}$ & $E3_{3}$ \\
  $E_{L-Aa}$ & -88.47 & -82.82 & -82.91 & -82.43 & -79.37 & & -77.54 & -81.36 & -75.18 & -67.44 & -81.73 & & -101.02 & -101.44 & -98.73 \\
  $E_{L-Aa(BS)}$ & -78.85 & -75.73 & -73.86 & -74.14 & -71.84 & & -74.85 & -77.92 & -66.57 & -61.33 & -76.75 & & -95.67 & -90.26 & -85.86 \\  
  $E_{L-w(BS)}$ & -3.21 & -0.70 & -6.90 & -1.50 & -7.70 & & & -8.69 & -6.62 & -11.43 & 1.45 & & 1.16 & -6.18 & -8.66 \\
  $E_{int(BS)}^{RS(a)}$ & -82.06 & -76.43 & -80.76 & -75.64 & -79.54 & & -74.85 & -86.61 & -73.19 & -72.76 & -75.30 & & -94.51 & -96.44 & -94.52 \\    
  $E_{int}^{RS(b)}$ & -91.68 & -83.52 & -89.81 & -83.93 & -87.07 & & -77.54 & -90.05 & -81.80 & -78.87 & -80.28 & & -99.86 & -107.62 & -107.39 \\ 
  $E_{int(BS)}^{(c)}$ & & & -79.75 & & & & & & -75.09 & & & & & -94.55 & \\    
  $E_{int}^{(d)}$ & & & -88.52 & & & & & & -78.73 & & & & & -100.11 & \\
  \toprule[1pt]
  & \multicolumn{5}{c}{{GNT-ER-w}}
  & \multicolumn{7}{c}{{DES-ER-w}}
  & \multicolumn{3}{c}{{BPA-ER-w}} \\
  \midrule[1pt]
  \ \ \ RS & & $GNT_{1}$ & $GNT_{2}$ & $GNT_{3}$ & & & & $DES_{1}$ & $DES_{2}$ & $DES_{3}$ & & & $BPA_{1}$ & $BPA_{2}$ & $BPA_{3}$ \\
  $E_{L-Aa}$ & & -95.40 & -94.95 & -94.35 & & & & -81.15 & -82.01 & -85.27 & & & -85.55 & -66.98 & -84.83 \\
  $E_{L-Aa(BS)}$ & & -90.11 & -89.58 & -87.87 & & & & -75.00 & -77.58 & -80.36 & & & -76.80 & -61.17 & -78.03 \\  
  $E_{L-w(BS)}$ & & -0.30 & -1.40 & -0.90 & & & & -1.26 & -0.64 & -0.82 & & & -0.87 & -10.21 & -0.45 \\  
  $E_{int(BS)}^{RS}$ & & -90.41 & -90.98 & -88.77 & & & & -76.26 & -78.22 & -81.18 & & & -77.67 & -71.38 & -78.48 \\    
  $E_{int}^{RS}$ & & -95.70 & -96.35 & -95.25 & & & & -82.41 & -82.65 & -86.09 & & & -86.42 & -77.19 & -85.28 \\  
  $E_{int(BS)}$ & & & -90.44 & & & & & & -76.61 & & & & & -76.48 & \\    
  $E_{int}$ & & & -95.77 & & & & & & -82.57 & & & & & -84.66 & \\
  \toprule[1pt]
  & \multicolumn{5}{c}{{BPAF-ER-w}}
  & \multicolumn{7}{c}{{HPTE-ER-w}}
  & \multicolumn{3}{c}{{DMDT-ER-w}} \\
  \midrule[1pt]
  \ \ \ RS &  & $BPAF_{1}$ & $BPAF_{2}$ & $BPAF_{3}$ & & & & $HPTE_{1}$ & $HPTE_{2}$ & $HPTE_{3}$ & & & $DMDT_{1}$ & $DMDT_{2}$ & $DMDT_{3}$ \\
  $E_{L-Aa}$ & & -70.15 & -65.49 & -64.68 & & & & -78.33 & -74.08 & -74.60 & & & -41.17 & -43.85 & -40.75 \\
  $E_{L-Aa(BS)}$ & & -67.66 & -65.66 & -60.73 & & & & -72.13 & -77.16 & -74.55 & & & -44.92 & -45.20 & -41.66 \\  
  $E_{L-w(BS)}$ & & -10.50 & -12.79 & -17.24 & & & & 1.82 & 1.97 & 1.89 & & & -4.29 & -6.30 & -3.62 \\  
  $E_{int(BS)}^{RS}$ & & -78.16 & -78.45 & -77.97 & & & & -70.31 & -75.19 & -72.66 & & & -49.21 & -51.50 & -45.28 \\    
  $E_{int}^{RS}$ & & -80.65 & -78.28 & -81.92 & & & & -76.51 & -72.11 & -72.71 & & & -45.46 & -50.15 & -44.37 \\  
  $E_{int(BS)}$ & & & -78.16 & & & & & & -70.92 & & & & & -49.56 & \\      
  $E_{int}$ & & & -80.63 & & & & & & -75.93 & & & & & -46.32 & \\  
  \bottomrule[1pt]
\end{tabular}
\label{Table V}
\begin{tablenotes}
      \footnotesize
      \item $BS$: Binding Site $L$: Ligand; $Aa$: Amino acids; $w$: waters. $^{(a)}$ $E_{int(BS)}^{RS} = (E_{L-Aa(BS)} + E_{L-w(BS)})$. $^{(b)}$ $E_{int}^{RS} = (E_{L-Aa} + E_{L-w(BS)})$. $^{(c)}$ The ligand-receptor FMO interaction energy at the binding site and correspond to the average weighted energy based on the cluster population. $^{(d)}$ The whole ligand-receptor FMO interaction energy and correspond to the average weighted energy based on the cluster population. All energies at MP2/6-31G(d) level of theory in kcal/mol.
\end{tablenotes}
  \end{threeparttable}%
   }
    \end{center}
\end{table*}

Figure 3 shows the canonical binding site, Figures 4-6 show stereoviews of estrogen receptor binding sites for the selected ligands and Figure 7 show the L-ER FMO interaction energy vs distortion factor.

\subsection{FMO Interaction Energies}

To facilitate the study, it is possible to identify a pattern in the ligands consisting of three regions that represent the ER pharmacophore, first described in 1950 as two hydroxyl groups separated by a hydrophobic spacer \cite{57}. The regions are: (i) the phenol A-ring; (ii) the
central hydrophobic core; (iii) the terminal hydroxylated ring. The exception is DMDT which contains methoxy substituents on the rings. An essential structural feature in most estrogen analogs is the presence of OH groups in the rings that enable the formation of hydrogen bonds at each end of the molecule. 

Table V shows the FMO interaction energy in the several systems. The order of the L-ER FMO interaction energies E$_{int}$ is as follows: \\

\begin{figure*}[!tb]
  \begin{center}
   \newcommand\tparbox[2]{\protect\parbox[t]{#1}{\protect\raggedright #2}}
    \subfigure[\tparbox{2.5cm}{17$\alpha E2_{1}$}]{
        \hbox{\hspace{-9.0em} \includegraphics[scale=0.24]{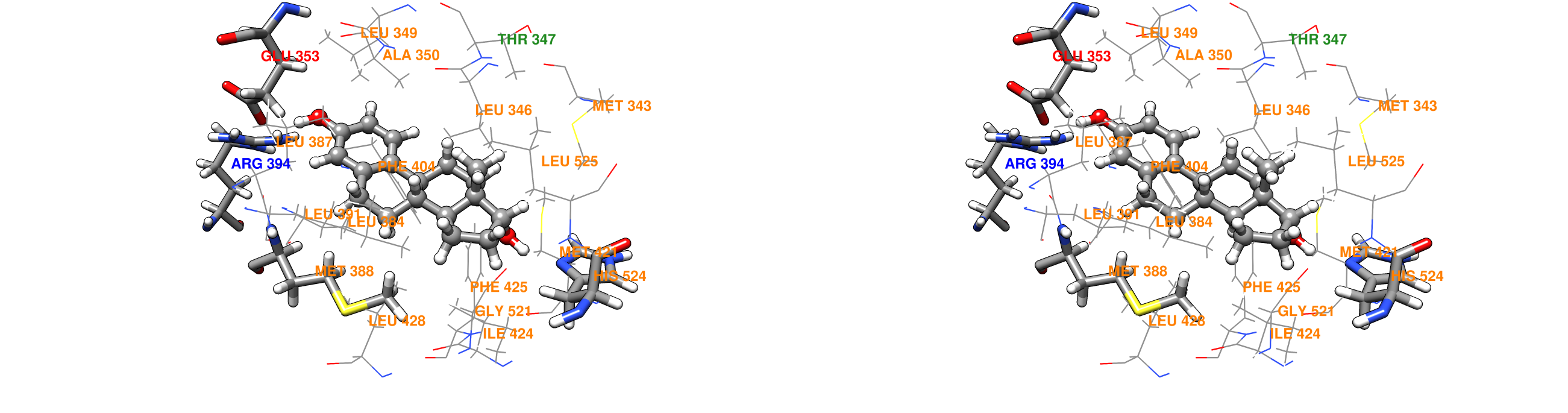}}
        }
    \subfigure[\tparbox{2.5cm}{17$\alpha E2_{5}$}]{
        \hbox{\hspace{-8.5em} \includegraphics[scale=0.24]{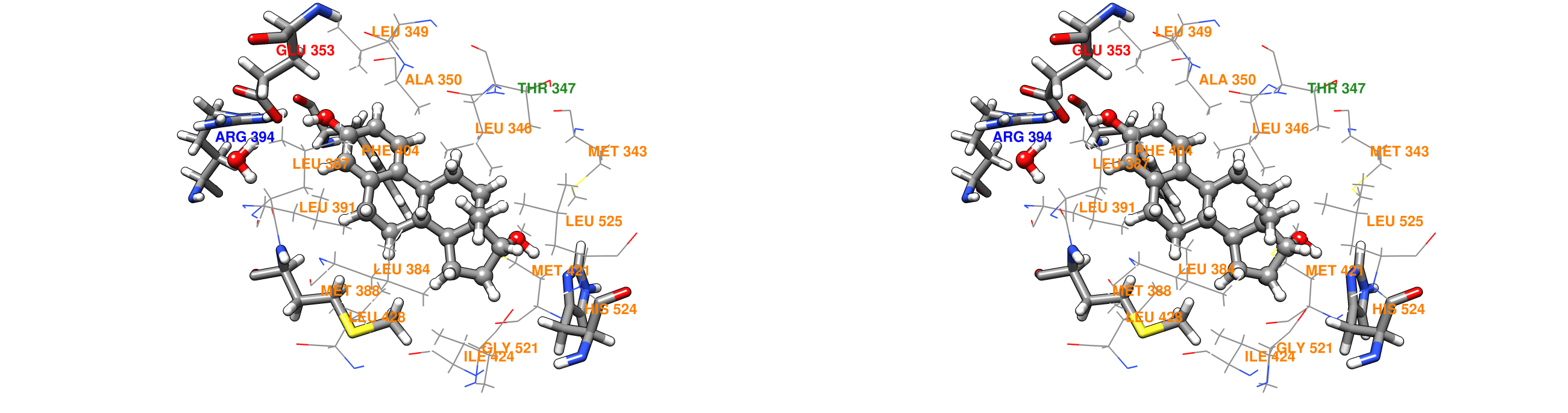}}
        }
        \subfigure[\tparbox{3.5cm}{$E3_{1}$}]{
        \hbox{\hspace{-7.5em} \includegraphics[scale=0.25]{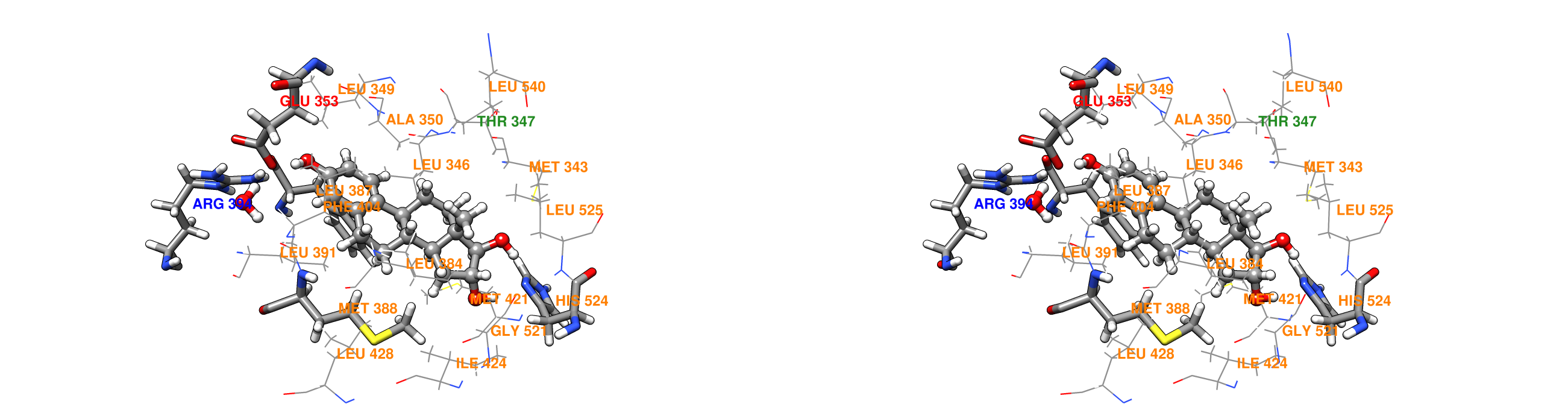}}
        }
    \caption{Stereo view of estrogen receptor binding site: ligand and water molecule (ball and stick); Glu 353, Met 388, Arg 394 and His 524 (stick); Phe 404 (stick or wire) and remaining amino acids residues (wire). Amino acid letter code: orange (apolar), green (polar), red (negatively charged), blue (positively charged).}
    \label{Figure 4} 
  \end{center}
\end{figure*}

\begin{figure*}[!hpt]
  \begin{center}
   \newcommand\tparbox[2]{\protect\parbox[t]{#1}{\protect\raggedright #2}}
    \subfigure[\tparbox{3.0cm}{$GNT_{1}$}]{
        \hbox{\hspace{-5.5em} \includegraphics[scale=0.24]{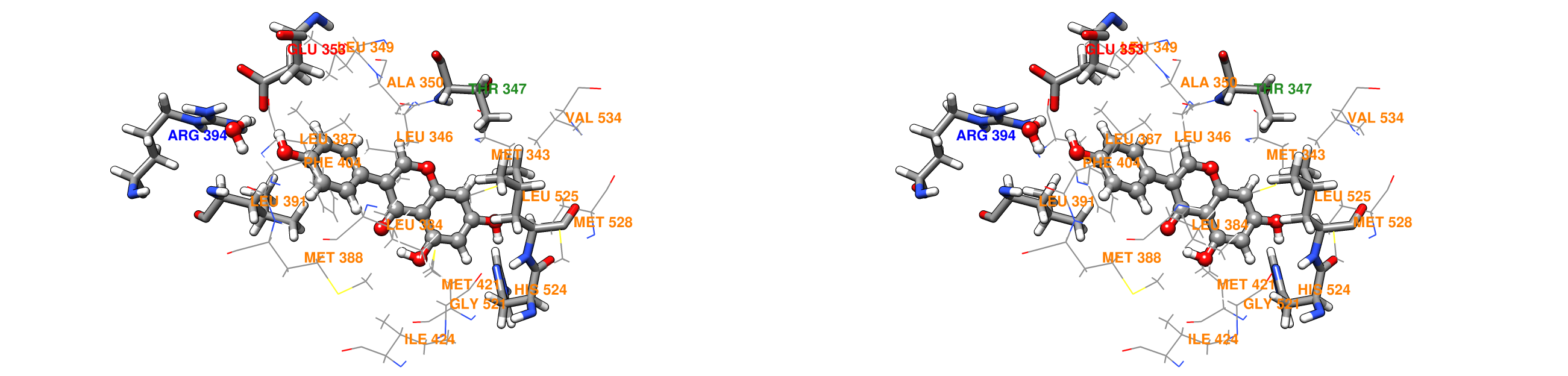}}
        }
    \subfigure[\tparbox{3.5cm}{$DES_{1}$}]{
        \hbox{\hspace{-6.0em} \includegraphics[scale=0.25]{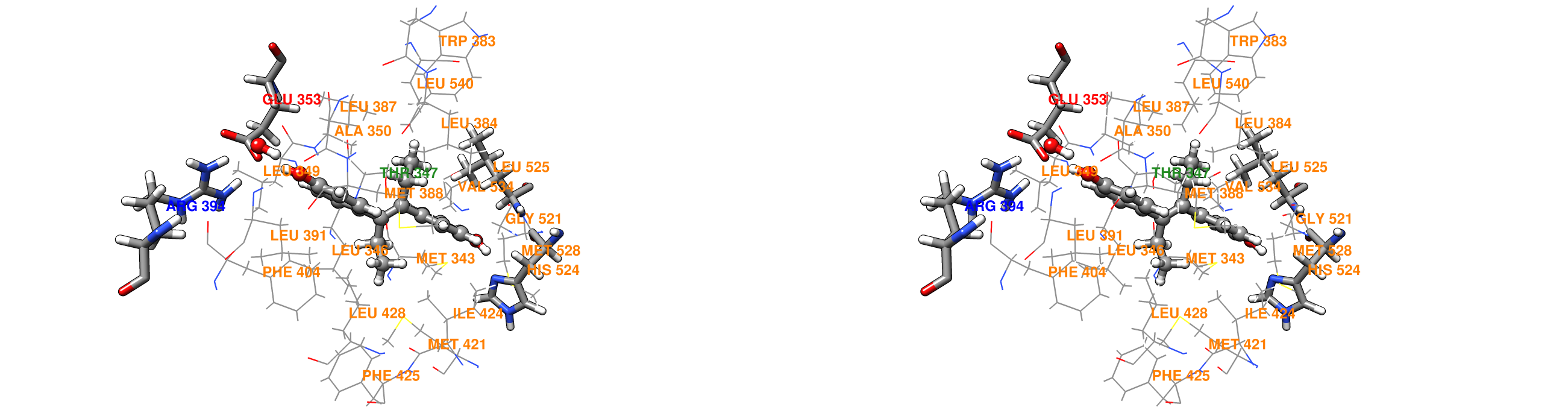}}
        }
    \subfigure[\tparbox{3.0cm}{$BPA_{1}$}]{
        \hbox{\hspace{-7.0em} \includegraphics[scale=0.25]{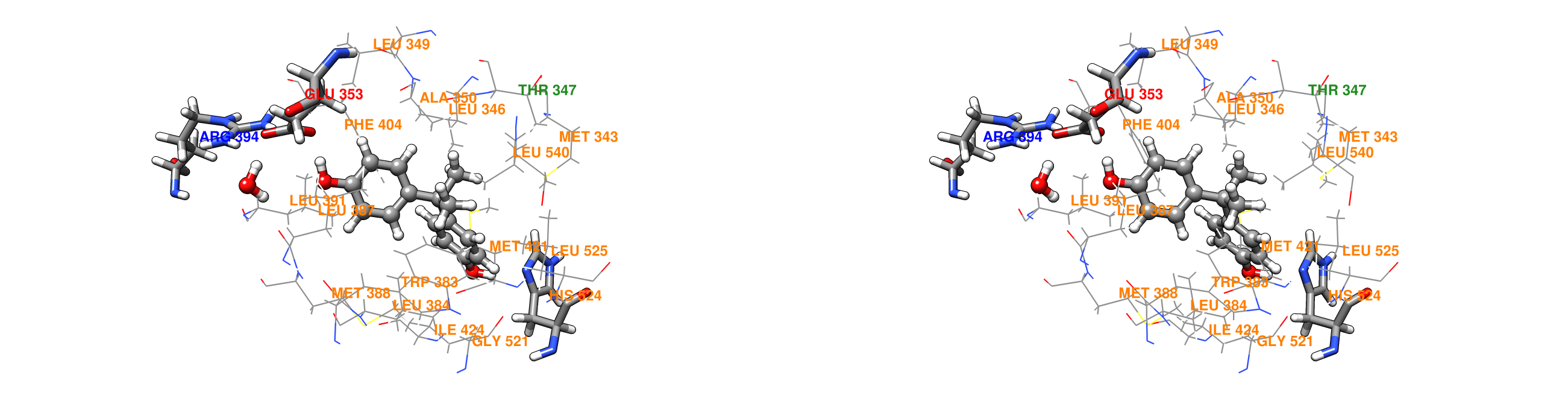}}
        }
    \caption{Stereo view of estrogen receptor binding site: ligand and water molecule (ball and stick); Glu 353, Arg 394 and His 524 (stick); Thr 347, Leu 391, Leu 525 (stick or wire) and remaining amino acids residues (wire). Amino acid letter code: orange (apolar), green (polar), red (negatively charged), blue (positively charged).}
    \label{Figure 5}
  \end{center}
\end{figure*}
\begin{figure*}[!tb]
  \begin{center}
   \newcommand\tparbox[2]{\protect\parbox[t]{#1}{\protect\raggedright #2}}  
    \subfigure[\tparbox{4.0cm}{$BPAF_{1}$}]{
        \hbox{\hspace{-7.5em} \includegraphics[scale=0.26]{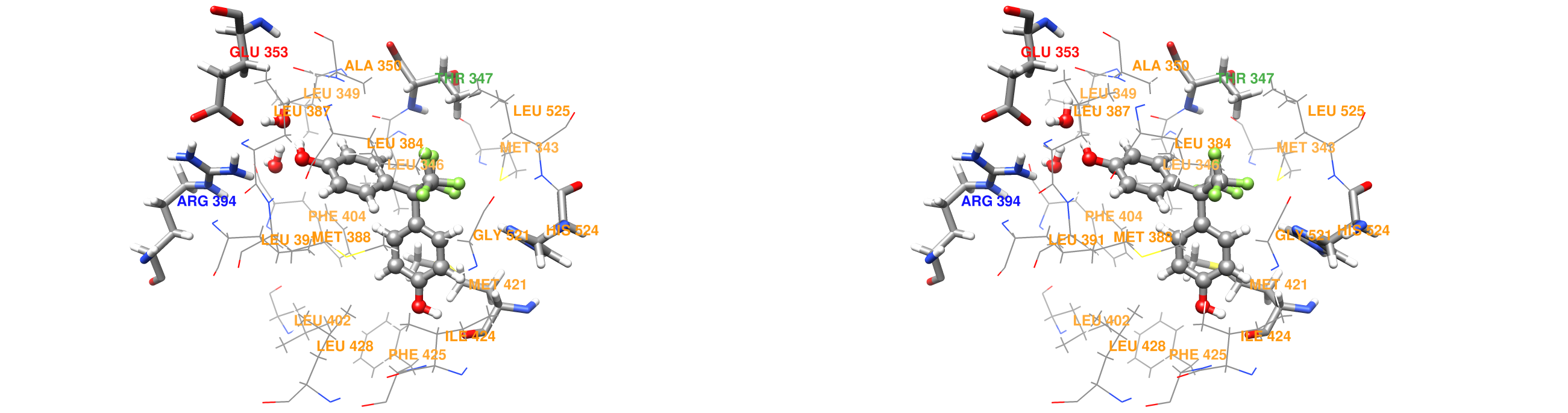}}
        }
    \subfigure[\tparbox{3.0cm}{$HPTE_{1}$}]{
        \hbox{\hspace{-7.5em} \includegraphics[scale=0.25]{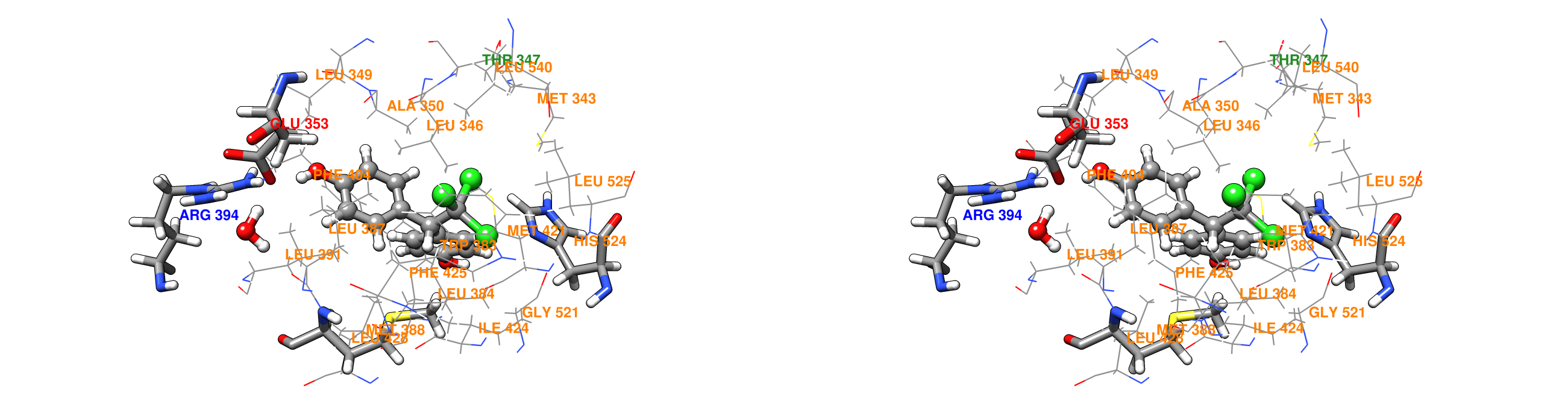}}
        }
    \subfigure[\tparbox{3.0cm}{$DMDT_{1}$}]{
        \hbox{\hspace{-7.5em} \includegraphics[scale=0.25]{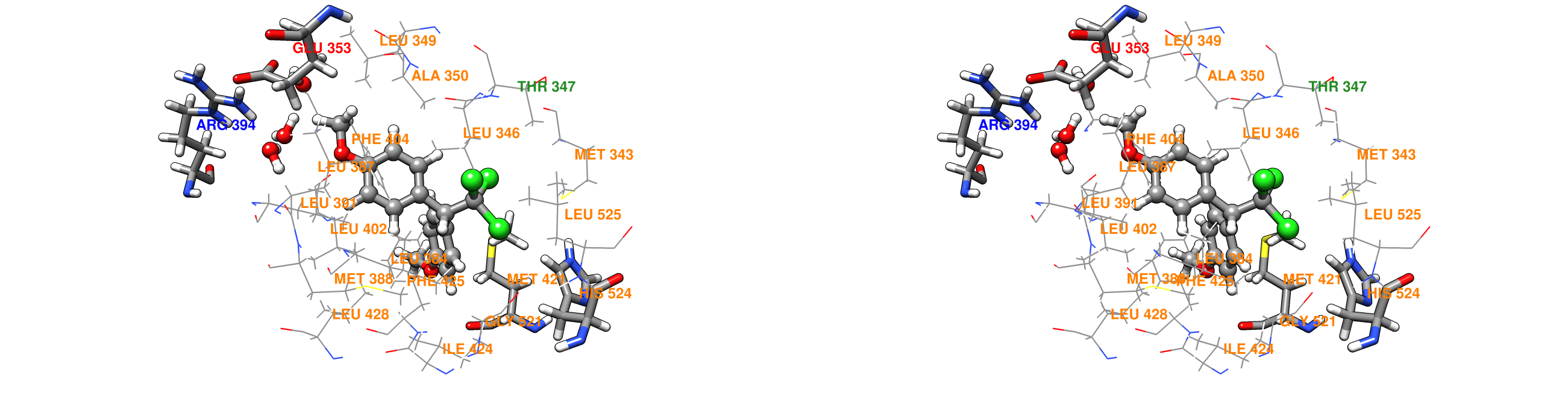}}
        }
    \caption{Stereo view of estrogen receptor binding site: ligand and water molecule (ball and stick); Glu 353, Arg 394 and His 524 (stick); Thr 347, Met 388, Met 421 (stick or wire) and remaining amino acids residues (wire). Amino acid letter code: orange (apolar), green (polar), red (negatively charged), blue (positively charged).}
    \label{Figure 6}
  \end{center}
\end{figure*}

\begin{table*}[!htbp]
 \begin{center}
  \resizebox{\textwidth}{!}{%
   \begin{threeparttable}
    \caption{FMO Ligand-Water Interaction Energy}
\begin{tabular}{ccccccccccccccccc}
  \toprule[1pt]  
  & \multicolumn{1}{c}{{$E2_{1}$}}
  & \multicolumn{1}{c}{{$E2_{2}$}}
  & \multicolumn{1}{c}{{$E2_{3}$}}
  & \multicolumn{1}{c}{{$E2_{4}$}}
  & \multicolumn{1}{c}{{$E2_{5}$}}
  & \multicolumn{1}{c}{{17$\alpha E2_{1}$}}
  & \multicolumn{1}{c}{{17$\alpha E2_{2}$}}
  & \multicolumn{1}{c}{{17$\alpha E2_{3}$}}
  & \multicolumn{1}{c}{{17$\alpha E2_{4}$}}
  & \multicolumn{1}{c}{{17$\alpha E2_{5}$}}
  & \multicolumn{1}{c}{{$E3_{1}$}}
  & \multicolumn{1}{c}{{$E3_{2}$}}
  & \multicolumn{1}{c}{{$E3_{3}$}}
  & \multicolumn{1}{c}{{$GNT_{1}$}}
  & \multicolumn{1}{c}{{$GNT_{2}$}}
  & \multicolumn{1}{c}{{$GNT_{3}$}} \\
  \midrule[1pt]
  Wat 1 & -3.21 & -0.70 & -6.90 & -1.50 & -0.83 &  & -1.30 & -6.62 & 1.68 & 1.45 & 1.16 & -6.18 & -8.18 & -0.30 & -1.40 & -0.90 \\
  Wat 2 & & & & & 1.05 & & -7.39 & & -13.11 & & & & -0.47 & & & \\
  Wat 3 & & & & & -7.93 & & & & & & & & & & & \\
  \midrule[1pt]  
  $E_{L-w(BS)}$ & -3.21 & -0.70 & -6.90 & -1.50 & -7.70 & & -8.69 & -6.62 & -11.43 & 1.45 & 1.16 & -6.18 & -8.66 & -0.30 & -1.40 & -0.90 \\      
  \bottomrule[1pt]
\end{tabular}
\label{Table VI}
\begin{tablenotes}
      \footnotesize
      \item All energies at MP2/6-31G(d) level of theory in kcal/mol.
\end{tablenotes}
  \end{threeparttable}%
   }
    \end{center}
\end{table*}

\begin{table*}[!htbp]
 \begin{center}
  \resizebox{\textwidth}{!}{%
   \begin{threeparttable}
    \caption{FMO Ligand-Water Interaction Energy (Cont.)}
\begin{tabular}{cccccccccccccccc}
  \toprule[1pt]  
  & \multicolumn{1}{c}{{$DES_{1}$}}
  & \multicolumn{1}{c}{{$DES_{2}$}}
  & \multicolumn{1}{c}{{$DES_{3}$}}
  & \multicolumn{1}{c}{{$BPA_{1}$}}
  & \multicolumn{1}{c}{{$BPA_{2}$}}
  & \multicolumn{1}{c}{{$BPA_{3}$}}
  & \multicolumn{1}{c}{{$BPAF_{1}$}}
  & \multicolumn{1}{c}{{$BPAF_{2}$}}
  & \multicolumn{1}{c}{{$BPAF_{3}$}}
  & \multicolumn{1}{c}{{$HPTE_{1}$}}
  & \multicolumn{1}{c}{{$HPTE_{2}$}}
  & \multicolumn{1}{c}{{$HPTE_{3}$}}
  & \multicolumn{1}{c}{{$DMDT_{1}$}}
  & \multicolumn{1}{c}{{$DMDT_{2}$}}
  & \multicolumn{1}{c}{{$DMDT_{3}$}} \\
  \midrule[1pt]
  Wat 1 & -1.26 & -0.64 & -0.82 & -0.87 & -10.21 & -0.45 & -12.18 & -12.79 & -12.56 & 1.82 & 1.97 & 1.89 & 0.66 & -6.23 & -1.35 \\
  Wat 2 & & & & & & & 1.68 & & -0.49 & & & & -1.82 & -0.07 & -2.35 \\
  Wat 3 & & & & & & & & & -4.19 & & & & -3.13 & & 0.08 \\
  \midrule[1pt]  
  $E_{L-w(BS)}$ & -1.26 & -0.64 & -0.82 & -0.87 & -10.21 & -0.45 & -10.50 & -12.79 & -17.24 & 1.82 & 1.97 & 1.89 & -4.29 & -6.30 & -3.62 \\      
  \bottomrule[1pt]
\end{tabular}
\label{Table VII}
\begin{tablenotes}
      \footnotesize
      \item All energies at MP2/6-31G(d) level of theory in kcal/mol.
\end{tablenotes}
  \end{threeparttable}%
   }
    \end{center}
\end{table*}

\begin{figure} [!htp]
\begin{center}
\hbox{\hspace{-2.5em} \includegraphics[scale=0.67]{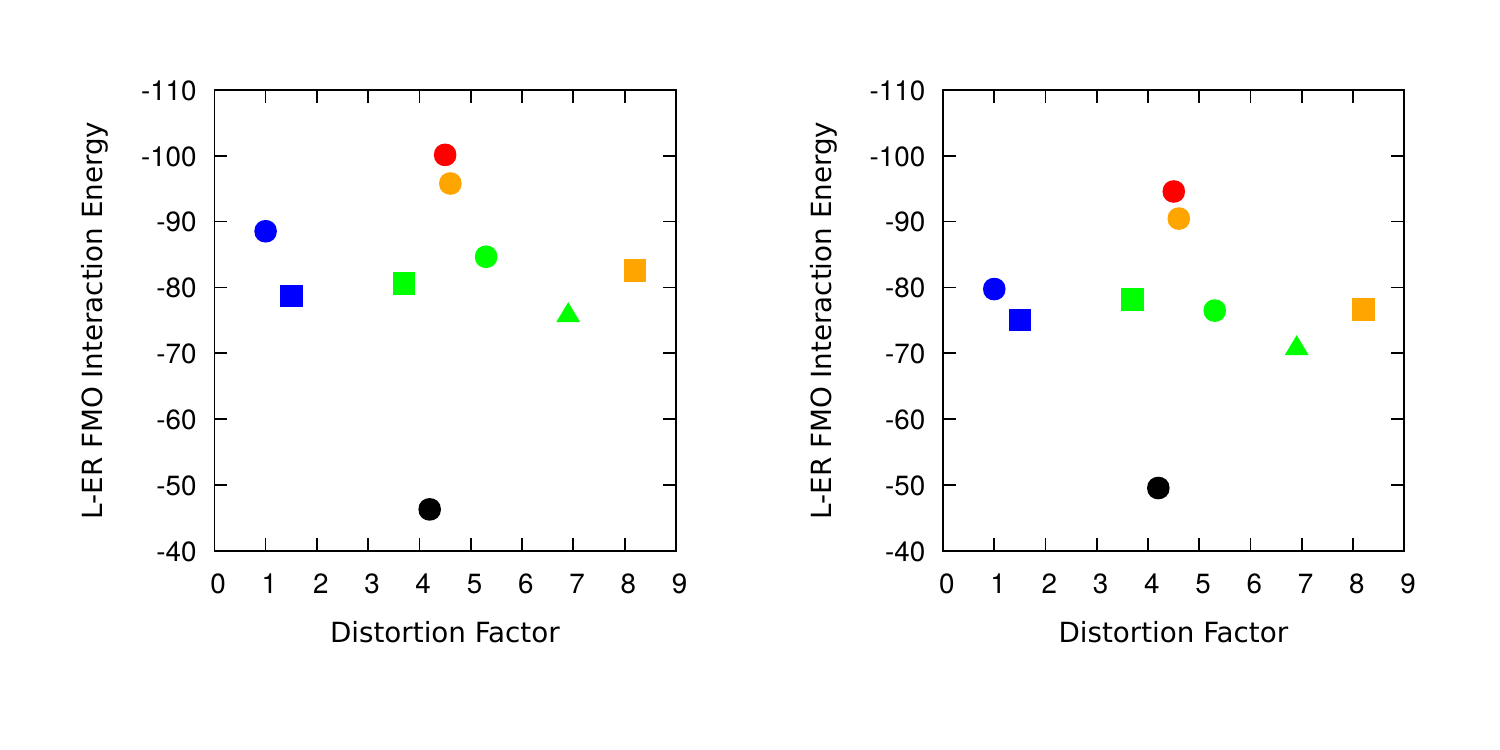}}
\end{center}
        \setlength{\abovecaptionskip}{-33pt}
    \caption{E$_{int}$ (Left) and E$_{int-(BS)}$ (Right) vs Distortion Factor. \hspace{0.6cm} 
    Blue: E2 (circle), 17$\alpha$-E2 (square); Red: (E3); Orange: GNT (circle), DES (square); Green: BPA (circle), BPAF (square), HPTE (triangle); Black: DMDT.}
    \label{Figure 7}
\end{figure}

\begin{figure} [!htp]
\begin{center}
\hbox{\hspace{-1.3em} \includegraphics[scale=0.23]{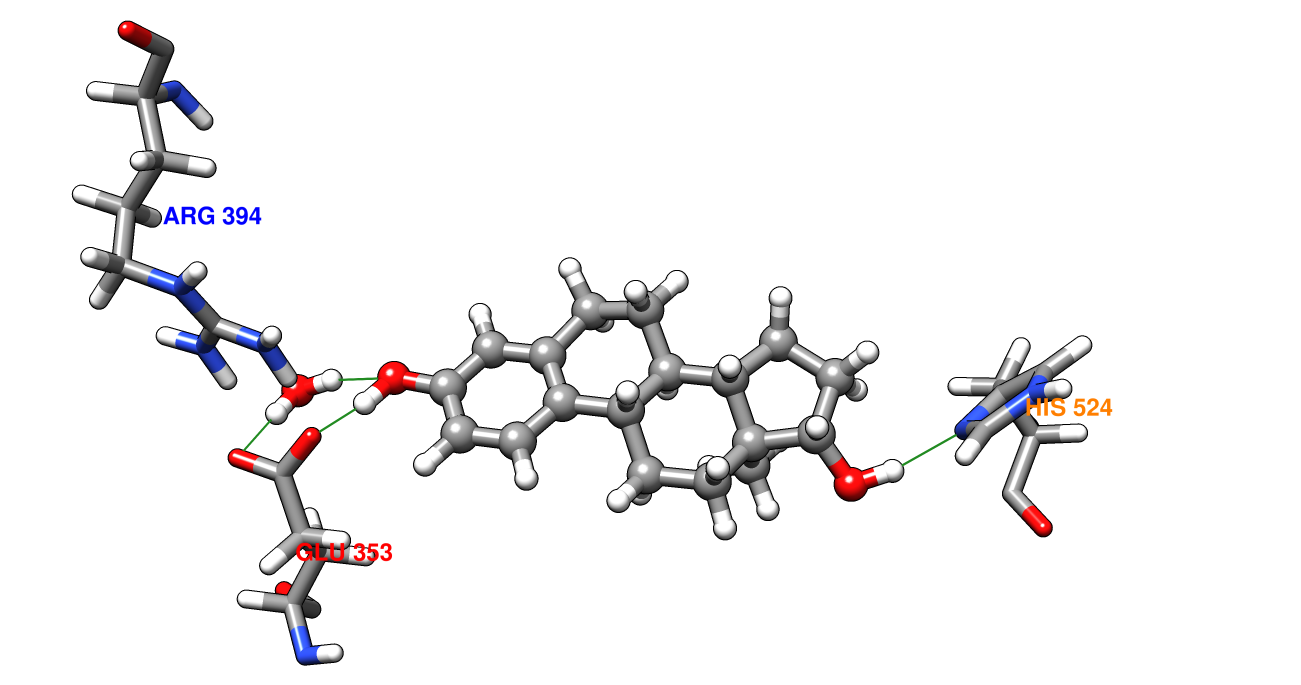}}
\end{center}
        \setlength{\abovecaptionskip}{-16pt}
    \caption{Intermolecular hydrogen bonds in the ligand binding site of E2$_{1}$. H-bonds are represented by forest green lines. His 524 is drawn as the $\epsilon$-tautomer.}
    \label{Figure 8}
\end{figure}

\begin{figure*}[!htp]
  \begin{center}
    \subfigure[17$\alpha E2_{5}$ \hspace{9cm} (d) $E3_{1}$]{
        \hbox{\hspace{-1.0em} \includegraphics[scale=0.35]{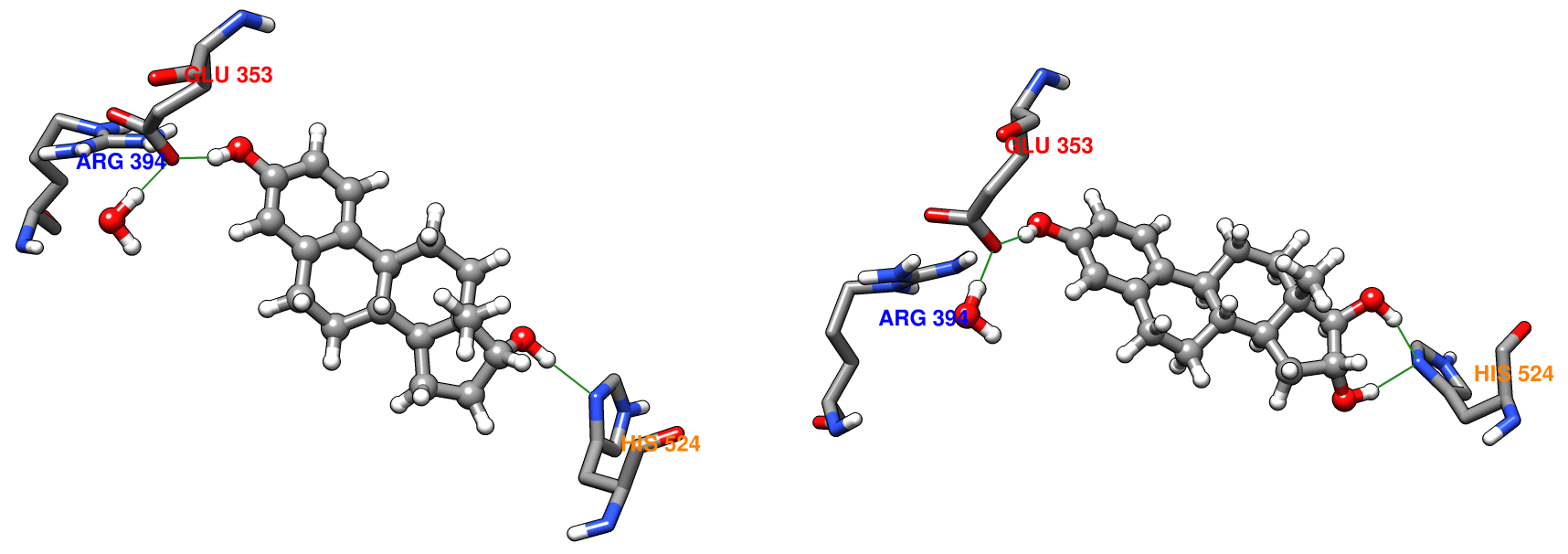}}
        }
    \subfigure[$DES_{1}$ \hspace{9cm} (e) $BPA_{1}$]{
        \hbox{\hspace{-1.0em} \includegraphics[scale=0.40]{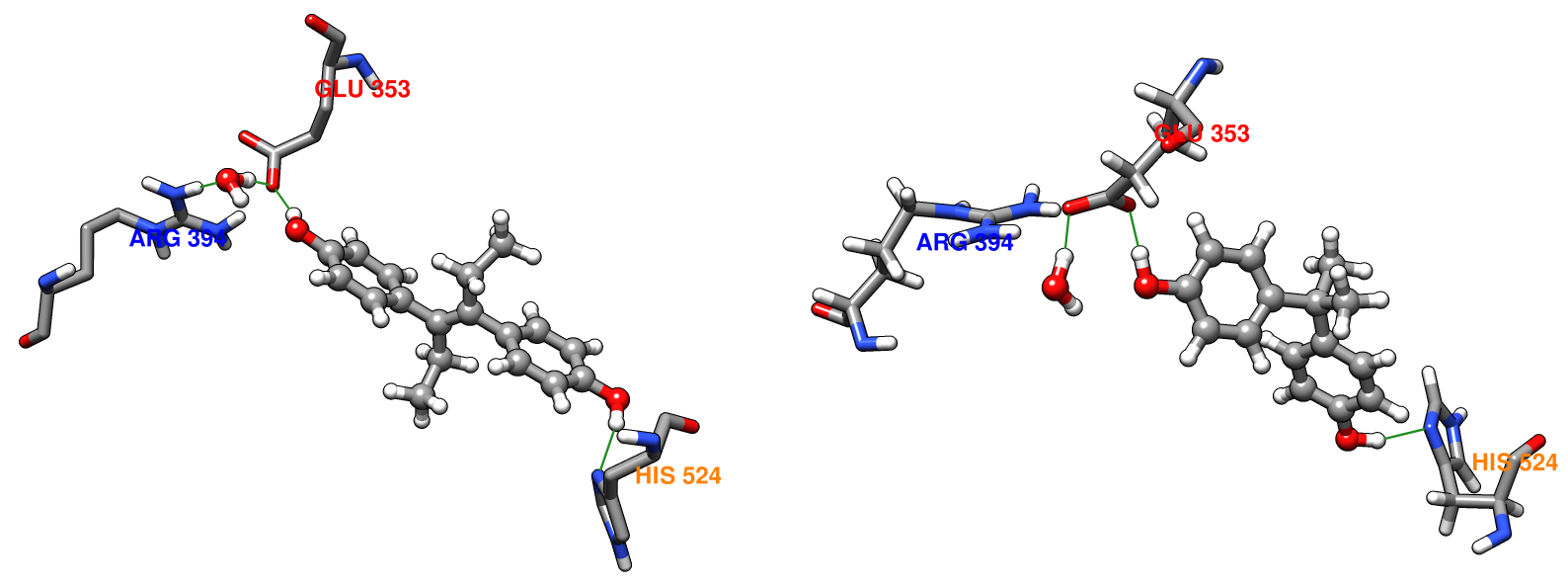}}
        }
    \subfigure[$BPAF_{1}$ \hspace{9cm} (f) $DMDT_{1}$]{
        \hbox{\hspace{-1.0em} \includegraphics[scale=0.40]{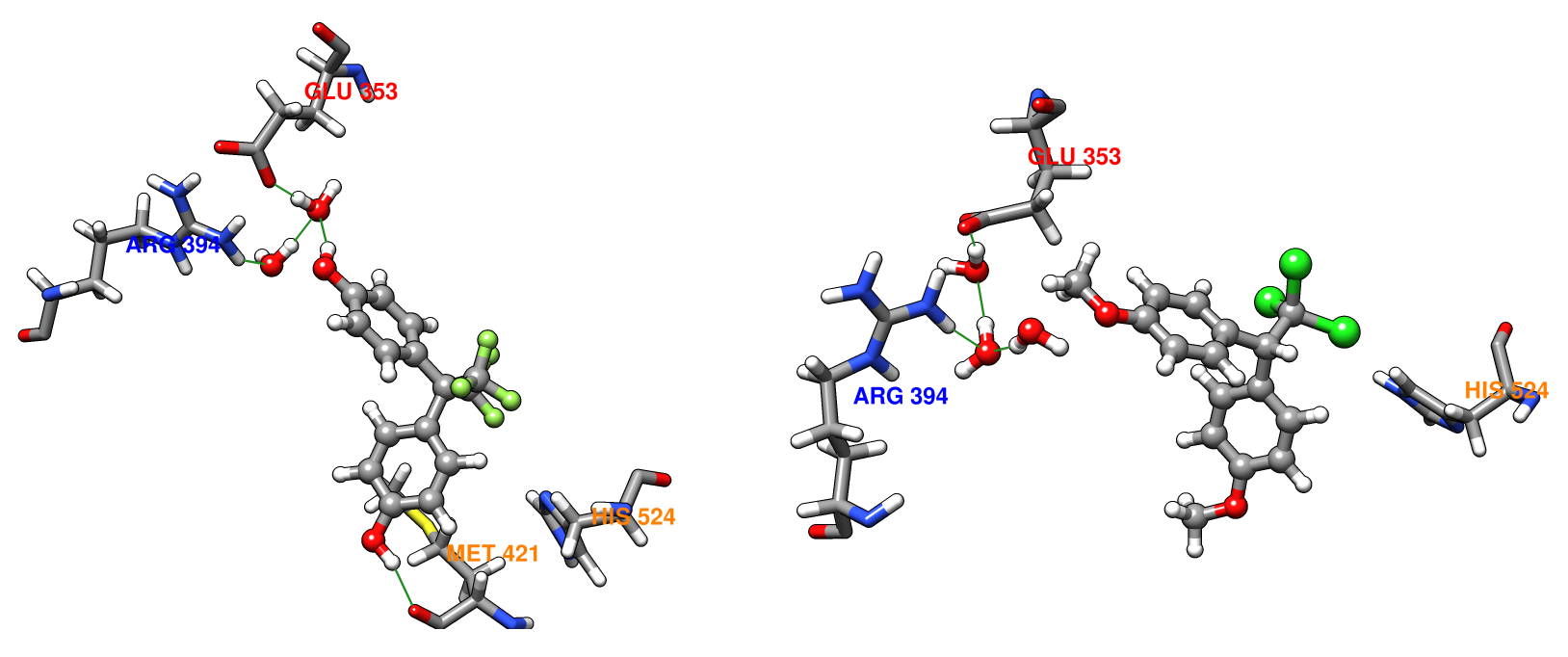}}
        }
    \caption{Intermolecular hydrogen bonds in the ligand binding site. The H-bonds configuration of 17$\alpha E2_{5}$ is found in GNT$_{1}$ RS, while that of DES$_{1}$ is found in HPTE$_{1}$ RS (except by ligand-His 524 hydrogen bond). H-bonds are represented by forest green lines. His 524 is drawn as the $\epsilon$-tautomer. }
    \label{Figure 9}
  \end{center}
\end{figure*}

\begin{scriptsize}
\scriptsize {E3 $<$ GNT $<$ E2 $<$ BPA $<$ DES $<$ BPAF $<$ 17$\alpha$-E2 $<$ HPTE $<$ DMDT}
\end{scriptsize}\\
\\
and was calculated as the weighted average energy based on the cluster population.

If we calculate the contribution of the binding site to E$_{int}$, we find that it is higher than 90\% in all cases. In the DMDT system it reaches a value of 107\%. This indicates that the choice of the interaction distance of 4.0 \r{A} is appropriate. In the particular case of DMDT we can conclude that most of the attractive interaction resides at the binding site.

The key L-ER interaction could be defined through Glu 353 and His 524 (Figures 8-9), since the interaction of these residues with the ligand generally contributes most to the interaction energy. Both residues are critical in anchoring the ligand to the binding site. The charged species Arg 394 and Glu 353 interacts very strongly (salt bridge) in all systems; E$_{int}$ in kcal/mol ranging from -109.6 ($DMDT_{1}$) to -130.4 ($BPAF)_{1}$. However, in most cases Arg 394 interacts repulsively with the ligand. It would seem that the function of Arg 394 is to stabilize Glu 353. 

Glu 353 interacts strongly with the ligand through a hydrogen bond with the hydroxyl group of region (i); donor-acceptor distance ranging from 2.52 ($E2_{1}$) to 2.64 (17$\alpha E2_{1}$ \r{A}). Nevertheless, there are two exceptions: in $BPAF_{1}$ the donor-acceptor distance is 3.51 \r{A} and the interaction is indirect through a water bridge (Figure 9c); in $DMDT_{1}$ the OH groups are replaced by methoxy groups, the distance between (Glu 353)-CO$_{2}^{-}$ and (DMDT)-OCH$_{3}$ is 5.18 \r{A} and probably the residue-ligand interaction is of the ion-dipole type (Figure 9f). No hydrogen bonds were found between the ligand and Arg 394; rather, the interaction is repulsive in nature, except in 17$\alpha E2_{4}$ which by population size is irrelevant. 

The presence or absence of water at the binding site affects the position and orientation of the amino acid residues and the ligand. Water molecules enable the formation of hydrogen bonds with the ligand and specific residues, usually Glu 353 and to a lesser extent Arg 394. Water bridges form between: two hydrogen bond acceptors, (Glu 353)-CO$_{2}^{-}$:$\cdot\cdot\cdot$H-O-H$\cdot\cdot\cdot$:OH-(E2) (Figure 8); a hydrogen bond acceptor and a hydrogen bond donor, (Glu 353)-CO$_{2}^{-}$:$\cdot\cdot\cdot$H-O:$\cdot\cdot\cdot$H$_{2}^{+}$N-(Arg 394) (where H-O is part of H-O-H) (Figure 9b). A higher order water bridge is found in $DMDT_{1}$, (Glu 353)-CO$_{2}^{-}$:$\cdot\cdot\cdot$H-O:$\cdot\cdot\cdot$H-O:$\cdot\cdot\cdot$H-O-H or (Glu 353)-CO$_{2}^{-}$:$\cdot\cdot\cdot$H-O:$\cdot\cdot\cdot$H-O:$\cdot\cdot\cdot$H$_{2}^{+}$N-(Arg 394) (Figure 9f). Intermolecular tricenter hydrogen bond centered on (Glu 353)-CO$_{2}^{-}$ and/or (His 524)-N$\delta$ is shown in Figures 9a, 9b and 9d. A complex network of hydrogen bond and double hydrogen bond is shown in Figure 9c and 9e, respectively. Weak ligand-water repulsive interaction were found in the following important RSs (Tables VI-VII): 17$\alpha E2_{5}$, $E3_{1}$ and $HPTE_{1}$. A strong attractive ligand-water interaction was found in $BPAF_{1}$ (-12.18 kcal/mol). Is the water present on the binding site required for a correct biological response?

His 524 interacts with the ligand through a hydrogen bond and like Glu 353 is the residue that contributes most to the interaction energy; donor-acceptor distance ranging from 2.74 ($GNT_{1}$) to 2.90 ($E3_{1}$ \r{A}). A notable exception is DMDT, which as we have seen, also do not form hydrogen bonds with Glu 353. The other exceptions are BPAF and HPTE. In the absence of the strong interaction with His 524, the anchoring of BPAF, HPTE and DMDT to the binding site is not adequate (see figure 6 and compare with figures 4-5). In these ligands, one of the aromatic rings tends to approach Glu 353 while the other moves away from His 524. Apparently, an interaction occurs between the sulfur of Met 421, Met 388 and the aromatic ring of EDCs, the so-called methionine-aromatic interaction \cite{58}: Met 421-aromatic (BPA, BPAF and DMDT RSs) and Met 388-aromatic (HPTE RS). The ligand-methionine interaction energies in kcal/mol for the most populated RSs are: $BPA_{1}$ (-4.66), $BPAF_{1}$ (-5.42), $DMDT_{1}$ (-6.01) and $HPTE_{1}$ (-8.98). Histidine-aromatic interactions \cite{59,60} could account for the attractive interaction energy in $BPAF_{1}$ (-6.26) and $HPTE_{1}$ (-6.58) kcal/mol. A strong hydrogen bond between the ligand and Met 421 was found in $BPAF_{1}$. From the above, it can be deduced that the ligand is anchored at the cost of significantly distorting the binding site. Compared to the other EDCs, BPA binding to ER by adopting a fashion similar to E2 with each of the phenol fragments pointing to to the ends of the binding site, one ring to residues Glu 353 and the other to His 524 (Figure 9e).

The central hydrophobic core of ligands interacts with several apolar residues of the ER, such as Met 343, Leu 346, Ala 350, Leu 387, Met 388, Leu 391, Phe 404, Met 421, Ile 424 and Leu 525 (Tables II-III). Although the interaction energies are not as large as those associated with anchor residues, their high number makes a difference. Also, the T-shape stacking between the aromatic group of ligands and the phenyl group of Phe 404 could still be favorable to the ligand binding (Figures 4b and 4c). The polar amino acid Thr 347 interacts with the ligand through the CH$_{3}$ group of the side chain and presents some interesting features; in EDCs the interaction is with the CX$_{3}$ group (X = H, F, Cl) and in the rest of the ligands with the rings, except in 17$\alpha E2_{1}$ where the interaction (weak) is with the methyl group. The strongest interactions occur in $GNT_{1}$ and $BPAF_{1}$. In the latter, it is likely that fluorine as an acceptor group interacts electrostatically with threonine through a pair of hydrogens of CH$_{3}$ group of the side chain \cite{61}.

Figure 7 (right) shows a relationship between the interaction energy of the ligand at the binding site (E$_{int(BS)}$) and the distortion factor. The relationship holds true for similar ligands: E2 and 17$\alpha$-E2; GNT and DES; BPA, BPAF and HPTE. Perhaps it is a coincidence, but it does highlight that there may be some relationship between the perturbation of the binding site and the interaction energy.

The experimental order of relative binding affinity (RBA) is as follows \cite{9,53,55,56}: \\

\begin{scriptsize}
\scriptsize {DES $>$ E2 $>$ E3 $>$ 17$\alpha$-E2 $>$ BPAF $>$ HPTE $>$ GNT $>$ BPA $>$ DMDT}
\end{scriptsize}\\

A theoretical study of the ligand-receptor interaction energy in aqueous media does not necessarily imply that we are able to predict the experimental ligand-receptor binding affinity. First, the binding energy is defined as the difference in energy between the fully optimized bound and unbound states: $$ E_{int} = E_{L-ER} - (E_{L} + E_{ER}) $$
where L and ER are considered non-interacting. The interaction energy is defined in the same way, except that it considers the ligand and the receptor in the geometry of the complex. Therefore, the difference between the both energies is the strain energy \cite{62}. Second and more importantly, the binding affinity, which is related to biological activity, is quantified by the association constant (Ka) of the ligand-receptor complex under standard state conditions and at the temperature considered, usually 298 K. This quantity expresses the equilibrium ratio of concentrations of the free receptor and ligand molecules to the bound complex. Consequently, the binding affinity is expressed as a thermodynamic equilibrium quantity related to the change in Gibbs free energy of the ligand-receptor complex formation. Thus, any interpretation or prediction of binding properties solely based on enthalpic or internal energy considerations must be inadequate, since the entropic contribution in $\Delta$G is not considered.

Each energy calculated in this work represents the interaction energy at a local minimum of the potential energy hypersurface of the particular ligand-receptor complex and is equivalent to the change in internal energy due to the formation of the complex at absolute zero, without considering the zero-point and strain energies. Nevertheless, the results of this class of studies are useful for insight into the factors involved in complex formation and therefore of relevance in drug design. In our model systems we have demonstrated what is generally observed in ligand-receptor complexes: the steric and chemical complementarity of the groups on the ligand and binding site surfaces \cite{63}. It was noted that apolar interactions are not directional and merely require atoms to approach each other, e.g. region (ii) of the ligand with the different apolar residues of the receptor. In contrast, hydrogen bonds are directional and chemical groups of the ligand and receptor must be properly oriented for binding to occur, e.g. regions (i) and (iii) with Glu 353 and His 524, respectively. Water molecules mediating an interaction between ligand and receptor (water bridges) were also observed. However, complex formation is not only determined by changes in ligand-receptor interactions as the ligand and receptor come together, but also by changes in receptor-water and ligand-water interactions, which in general will depend on: (i) the difference between the number of water molecules interacting with the ligand at the binding site and in the bulk solvent; (ii) the difference between the number of water molecules interacting with the receptor at the binding site before and after complex formation, e.g. in the most populated RS of the apo-receptor \cite{40}, eight water molecules were found in the binding site and in the presence of ligand this amount decreases considerably.

\section{Conclusions}

The molecular interactions between selected ligands and hER$\alpha$ LBD were calculated, and from these E$_{int}$ was obtained; the order in kcal/mol is as follows: \\

\begin{scriptsize}
\scriptsize {E3 (-100.1) $<$ GNT (-95.8) $<$ E2 (-88.5) $<$ BPA (-84.7) $<$ DES (-82.6) $<$ BPAF (-80.6) $<$ 17$\alpha$-E2 (-78.7) $<$ HPTE (-75.9) $<$ DMDT (-46.3)}
\end{scriptsize}\\
\\
and does not correspond to the experimental RBA order, mainly because the model does not consider thermal and entropic effects. The outlier ligand DMDT has the lowest RBA and the highest E$_{int}$ value (low affinity) as a result of the substitution of OH groups by methoxy groups in the rings, preventing the formation of hydrogen bonds with the anchor residues Glu 353 and His 524. E3 RSs present the lowest E$_{int}$ and it is justified because the ligand contains in the aromatic ring a pair of OH groups that interact quite well with His 524.

The concept of ligand-binding site complementarity is key to understanding the position and orientation of the ligand at the binding site. This "law" is fulfilled even at the cost of producing some binding site perturbation. For example, in BPAF1 the anchoring ligand His 524 is replaced by Met 421, leading to a distortion of the binding site. In an attempt to quantify this perturbation a factor was derived based on the missing and extra residues, obtaining a certain relationship between L-ER FMO interaction energy and this distortion factor. With regard to missing residues, Met 528 is not conserved in most of the RSs. It follows from the above that the binding site is not a totally unalterable entity in terms of the residues that define it. This behavior is true not only in relation to different ligands, but to the same ligand at different times in a molecular dynamics simulation. This is probably also true in real systems.

The central hydrophobic core of the ligands interacts attractively with several apolar residues; these interactions play an important role in stabilizing the ligands at the binding site. Glu 353 and His 524 interacts strongly with most ligands through a hydrogen bond with the hydroxyl group of region (i) and region (iii), respectively. These residues are essential in anchoring the ligand to the binding site. Water molecules were found at the binding site of all representative structures, except 17$\alpha E2_{1}$. Other types of interactions observed in RSs are: salt bridges between Arg 394 and Glu 353, water bridges, methionine-aromatic interactions between the sulfur of Met 421 or Met 388 and the aromatic ring of EDCs, histidine-aromatic interactions in $BPAF_{1}$ and $HPTE_{1}$, T-shape stacking between the aromatic ring of ligands and the phenyl group of Phe 404 and Thr 347-ligand interactions. It should be noted that Arg 394 interacts repulsively with the ligands and although not demonstrated in this study, it is possibly a key residue in the ligand dissociation process.

\section*{Acknowledgements}

I wish to express my gratitude to arXiv repository and to each and every person who has contributed to the development and maintenance of free and open source software.

\bibliographystyle{ieeetr}
\bibliography{manuscrito}

\begin{thebibliography}{63}

\bibitem{1}
E.K. Shanle, W. Xu, Endocrine Disrupting Chemicals Targeting Estrogen Receptor
Signaling: Identification and Mechanisms of Action,
\newblock {\em Chem. Res. Toxicol.} 24 (2011) 6–19. https://doi.org/10.1021/tx100231n.

\bibitem{2}
A.M. Brzozowski, A.C.W. Pike, Z. Dauter, R.E. Hubbard, T. Bonn, O. Engstrom, L. Ohman † , G.L. Greene, J-Å. Gustafsson, M. Carlquist, Molecular basis of agonism and antagonism in the oestrogen receptor,
\newblock {\em Nature} 389 (1997) 293–342. https://doi.org/10.1038/39645.

\bibitem{3}
E. Diamanti-Kandarakis, J-P. Bourguignon, L.C. Giudice, R. Hauser, G.S. Prins, A.M. Soto, R.T. Zoeller, A.C. Gore, Endocrine-Disrupting Chemicals: An Endocrine Society Scientific Statement,
\newblock {\em Endocrine Reviews} 30(4) (2009) 753–758. https://doi.org/10.1210/er.2009-0002.

\bibitem{4}
S. Sasson, A.C. Notides, Estriol and Estrone Interaction with the Estrogen Receptor. II. Estriol and estrone-induced inhibition of the cooperative binding of [3H]estradiol to the estrogen receptor, 
\newblock {\em J. Biol. Chem.} 258 (1983) 8118-8122.

\bibitem{5}
B.S. Katzenellenbogen, Biology and receptor interactions of estriol and estriol derivatives in vitro and in vivo, 
\newblock {\em J. Steroid. Biochem.} 20 (1984) 1033-1037.

\bibitem{6}
G.G.J.M. Kuiper, J.G. Lemmen, B. Carlsson, J.C. Corton, S.H. Safe, P.T. van der Saag, B. van der Burg, J-A. Gustafsson, Interaction of Estrogenic Chemicals and Phytoestrogens with Estrogen Receptor $\beta$,
\newblock {\em Endocrinology} 139 (1998) 4252-4263. https://doi.org/10.1210/endo.139.10.6216.

\bibitem{7}
K.W. Gaido, S.C. Maness, D.P. McDonnell, S.S. Dehal, D. Kupfer, S. Safe, Interaction of Methoxychlor and Related Compounds with Estrogen Receptor $\alpha$ and $\beta$, and Androgen Receptor: Structure-Activity Studies,
\newblock {\em Mol. Pharmacol.} 58 (2000) 852-858. https://doi.org/10.1124/mol.58.4.852.

\bibitem{8}
A. Tamrazi, K.E. Carlson, J.A. Katzenellenbogen, Molecular Sensors of Estrogen Receptor Conformations and Dynamics,
\newblock {\em Mol. Endocrinol.} 17 (2003) 2593-602. https://doi.org/10.1210/me.2003-0239.

\bibitem{9}
A. Matsushima, X. Liu, H. Okada, M. Shimohigashi, Y. Shimohigashi, Bisphenol AF Is a Full Agonist for the Estrogen Receptor ER$\alpha$ but a Highly Specific Antagonist for ER$\beta$,
\newblock {\em Environmental Health Perspectives} 118 (2010) 1267-1272.  https://dx.doi.org/10.1289

\bibitem{10}
C.N. Harvey, J.C. Chen, C.A. Bagnell, M. Uzumcu, Methoxychlor and Its Metabolite HPTE Inhibit cAMP Production and Expression of Estrogen Receptors $\alpha$ and $\beta$ in the Rat Granulosa Cell In Vitro,
\newblock {\em Reprod Toxicol.}  51 (2015) 72-78. https://dx.doi.org/10.1016

\bibitem{11}
H. Cao, F. Wang, Y. Liang, H. Wang, A. Zhang, M. Song, Experimental and computational insights on the recognition mechanism between the estrogen receptor $\alpha$ with bisphenol compounds,
\newblock {\em Arch. Toxicol.} 91 (2017) 3897-3912. https://doi.org/10.1007/s00204-017-2011-0.

\bibitem{12}
R. Mesnage, A. Phedonos, M. Biserni, M. Arno, S. Balu, J.C. Corton, R. Ugarte, M.N. Antoniou, Evaluation of estrogen receptor alpha activation by glyphosate-based herbicide constituents,
\newblock {\em Food and Chem. Toxicol.} 108 (2017) 30-42. https://doi.org/10.1016/j.fct.2017.07.025.

\bibitem{13}
M.M.H. van Lipzig, A.M. ter Laak, A. Jongejan, N.P.E. Vermeulen, M. Wamelink, D. Geerke, J.H.N. Meerman, Prediction of Ligand Binding Affinity and Orientation of Xenoestrogens to the Estrogen Receptor by Molecular Dynamics Simulations and the Linear Interaction Energy Method,
\newblock {\em J. Med. Chem.} 47 (2004) 1018-1030. https://doi.org/10.1021/jm0309607.

\bibitem{14}
K. Fukuzawa, K. Kitaura, M. Uebayasi, K. Nakata, T. Kaminuma, T. Nakano, Ab initio Quantum Mechanical Study of the Binding Energies of Human Estrogen Receptor $\alpha$ with Its Ligands: An Application of Fragment Molecular Orbital Method,
\newblock {\em J. Comput. Chem.} 26 (2005) 1-10. https://doi.org/10.1002/jcc.20130.

\bibitem{15}
L. Celik, J.D.D. Lund, B. Schiøtt, Conformational Dynamics of the Estrogen Receptor $\alpha$: Molecular Dynamics Simulations of the Influence of Binding Site Structure on Protein Dynamics,
\newblock {\em Biochemistry} 46 (2007) 1743-1758. https://doi.org/10.1021/bi061656t.

\bibitem{16}
M.T. Sonoda, L. Martinez, P. Webb, M.S. Skaf, I. Polikarpov, Ligand Dissociation from Estrogen Receptor Is Mediated by Receptor Dimerization: Evidence from Molecular Dynamics Simulations,
\newblock {\em Mol. Endocrinol.} 22 (2008) 1565–1578. https://dx.doi.org/10.1210

\bibitem{17}
T.D. McGee, J. Edwards, A.E. Roitberg, Preliminary Molecular Dynamic Simulations of the Estrogen Receptor Alpha Ligand Binding Domain from Antagonist to Apo, Int. J. 
\newblock {\em Environ. Res. Public. Health} 5 (2008) 111-114.

\bibitem{18}
F. Spyrakis, P. Cozzini, How Computational Methods Try to Disclose the Estrogen Receptor Secrecy-Modeling the Flexibility,
\newblock {\em Curr. Med. Chem.} 16 (2009) 2987-3027. https://doi.org/10.2174/092986709788803123.

\bibitem{19}
S. Zhuang, J. Zhang, Y. Wen, C. Zhang, W. Liu, Distinct mechanisms of endocrine disruption of DDT-related pesticides toward estrogen receptor $\alpha$ and estrogen-related receptor $\gamma$,
\newblock {\em Environ. Toxicol. Chem.} 31 (2012) 2597-605. https://doi.org/10.1002/etc.1986.

\bibitem{20}
L. Gao, Y. Tu, H. Ågren, L.A. Eriksson, Characterization of Agonist Binding to His524 in the Estrogen Receptor $\alpha$ Ligand Binding Domain,
\newblock {\em J. Phys. Chem. B} 116 (2012) 4823-4830. https://doi.org/10.1021/jp300895g.

\bibitem{21}
D. Jereva, F. Fratev, I. Tsakovska, P. Alov, T. Pencheva, I. Pajeva, Molecular dynamics simulation of the human estrogen receptor alpha: contribution to the pharmacophore of the agonists,
\newblock {\em Math. Comput. Simulation} (2015). http://dx.doi.org/10.1016/j.matcom.2015.07.003.

\bibitem{22}
L. Li, Q. Wang, Y. Zhang, Y. Niu, X. Yao, H. Liu, 2015. The Molecular Mechanism of Bisphenol A (BPA) as an Endocrine Disruptor by Interacting with Nuclear Receptors: Insights from Molecular Dynamics (MD) Simulations.
\newblock {\em PLoS One.} 10, e0120330. https://dx.doi.org/10.1371

\bibitem{23}
A. Zafar, S. Ahmadb, I. Naseem, 2015. Insight into the structural stability of Coumestrol with Human Estrogen Receptor $\alpha$ and $\beta$ subtypes: A combined approach involving docking and molecular dynamics simulation studies.
\newblock {\em Wiley RSC Adv.} 5, 81295. https://doi.org/10.1039/C5RA14745J.

\bibitem{24}
A. Shtaiwi, R. Adnan, M. Khairuddean, M. Al-Qattan, 2018. Molecular dynamics simulation of human estrogen receptor free and bound to morpholine ether benzophenone inhibitor.
\newblock {\em Theor. Chem. Acc.} 137, 101. https://doi.org/10.1007/s00214-018-2277-1.

\bibitem{25}
M. Pavlin, A. Spinello, M. Pennati, N. Zaffaroni, S. Gobbi, A. Bisi, G. Colombo, A. Magistrato, 2018. A Computational Assay of Estrogen Receptor $\alpha$ Antagonists Reveals the Key Common Structural Traits of Drugs Effectively Fighting Refractory Breast Cancers.
\newblock {\em Sci. Rep.} 8, 649. https://doi.org/10.1038/s41598-017-17364-4.

\bibitem{26}
E.E.M. Eid, F. Azam, M. Hassan, I. M. Taban, M.A. Halim, Zerumbone binding to estrogen receptors: an in-silico investigation,
\newblock {\em Journal of Receptors and Signal Transduction} 38 (2018) 342-351. https://doi.org/10.1080/10799893.2018.1531886.

\bibitem{27}
H. Cao, L. Wang, M. Cao, T. Ye, Y. Sun, Computational insights on agonist and antagonist mechanisms of estrogen receptor $\alpha$ induced by bisphenol A analogues,
\newblock {\em Environmental Pollution} 248 (2019) 536-545. https://doi.org/10.1016/j.envpol.2019.02.058.

\bibitem{28}
E. Rossini, E. Giacopuzzi, F. Gangemi, M. Tamburello, D. Cosentini, A. Abate, M. Laganà, A. Berruti, S. Grisanti, S. Sigala, Estrogen-Like Effect of Mitotane Explained by Its Agonist Activity on Estrogen Receptor-$\alpha$,
\newblock {\em Biomedicines} 9(6) (2021) 681. https://doi.org/10.3390/biomedicines9060681.

\bibitem{29}
C. Kalaiarasi, S. Manjula, P. Kumaradhas, Combined quantum mechanics/molecular mechanics (QM/MM) methods to understand the charge density distribution of estrogens in the active site of estrogen receptors,
\newblock {\em RSC Adv.} 9 (2019) 40758–40771. http://doi.org/10.1039/c9ra08607b.

\bibitem{30}
T. Wang, Y. Wang, X. Zhuang, F. Luan, C. Zhao, M.N.D.S. Cordeiro, Interaction of Coumarin Phytoestrogens with ER $\alpha$ and ER $\beta$ : A Molecular Dynamics Simulation Study,
\newblock {\em Molecules} 25 (2020) 1165. http://doi.org/10.3390/molecules25051165.

\bibitem{31}
H. Tan, X. Wang, H. Hong, E. Benfenati, J.P. Giesy, G.C. Gini, R. Kusko, X. Zhang, H. Yu, W. Shi, Structures of Endocrine-Disrupting Chemicals Determine Binding to and Activation of the Estrogen Receptor $\alpha$ and Androgen Receptor,
\newblock {\em Environ. Sci. Technol.} 54 (2020) 11424-11433. https://dx.doi.org/10.1021/acs.est.0c02639.

\bibitem{32}
K. Kato, K. Fujii, T. Nakayoshi, Y. Watanabe, S. Fukuyoshi, K. Ohta, Y. Endo, N. Yamaotsu, S. Hirono, E. Kurimoto, A. Oda, 2018. Structural differences between the ligand-binding pockets of estrogen receptors alpha and beta.
\newblock {\em J. Phys.: Conf. Ser.} 1136, 012021. https://doi.org/10.1088/1742-6596/1136/1/012021.

\bibitem{33}
S. Lee, M.G. Barron, 2017. Structure-Based Understanding of Binding Affinity and Mode of Estrogen Receptor $\alpha$ Agonists and Antagonists.
\newblock {\em PLoS One.} 12, e0169607. https://dx.doi.org/10.1371

\bibitem{34}
T. Vreven, K. Morokuma, Hybrid Methods: ONIOM(QM:MM) and QM/MM,
\newblock {\em Annual Reports in Computational Chemistry} 2 (2006) 35-51. https://doi.org/10.1016/S1574-1400(06)02003-2.

\bibitem{35}
A.W. Gotz, M.A. Clark, R.C. Walker, An Extensible Interface for QM/MM Molecular Dynamics Simulations with AMBER,
\newblock {\em Journal of Computational Chemistry} 35 (2014) 95-108. https://doi.org/10.1002/jcc.23444.

\bibitem{36}
M.S. Gordon, D.G. Fedorov, S.R. Pruitt, L.V. Slipchenko, Fragmentation Methods: A Route to Accurate Calculations on Large Systems,
\newblock {\em Chem. Rev.} 112 (2012) 1632-672. https://doi.org/10.1021/cr200093j.

\bibitem{37}
D. Fedorov, K. Kitaura, Theoretical Background of the Fragment Molecular Orbital (FMO) Method and Its Implementation in GAMESS, in: D. Fedorov, K. Kitaura (Eds.), Fragment Molecular Orbital Method, CRC Press, Boca Raton, FL, 2009, pp. 5-36.

\bibitem{38}
D. Fedorov, K. Kitaura, Extending the Power of Quantum Chemistry to Large Systems with the Fragment Molecular Orbital Method,
\newblock {\em J. Phys. Chem. A} 111 (2007) 6904-6914. https://doi.org/10.1021/jp0716740.

\bibitem{39}
M.P. Mazanetz, E. Chudyk, D.G. Fedorov, Y. Alexeev, Applications of the Fragment Molecular Orbital Method to Drug Research, in: W. Zhang (Eds.), Computer-Aided Drug Discovery. Methods in Pharmacology and Toxicology. Humana Press, New York, NY, 2015.

\bibitem{40}
R. Ugarte, FMO Interaction Energy between 17$\beta$-Estradiol, 17$\alpha$-Estradiol and Human Estrogen Receptor $\alpha$,
\newblock {\em arXiv:2012.10822 [q-bio.BM]} (2020). https://arxiv.org/abs/2012.10822.

\bibitem{41}
S. Eiler, M. Gangloff, S. Duclaud, D. Moras, M. Ruff, Overexpression, Purification, and Crystal Structure of Native ER$\alpha$ LBD,
\newblock {\em Protein Expression and Purification} 22 (2001) 165-173. https://doi.org/10.1006/prep.2001.1409.

\bibitem{42}
D.A. Case, J.T. Berryman, R.M. Betz, D.S. Cerutti, T.E. Cheatham III, T.A. Darden, R.E. Duke, T.J. Giese, H. Gohlke, A.W. Goetz, N. Homeyer, S. Izadi, P. Janowski, J. Kaus, A. Kovalenko, T.S. Lee, S. LeGrand, P. Li, T. Luchko, R. Luo, B. Madej, K.M. Merz, G. Monard, P. Needham, H. Nguyen, H.T. Nguyen, I. Omelyan, A. Onufriev, D.R. Roe, A. Roitberg, R. Salomon-Ferrer, C.L. Simmerling, W. Smith, J. Swails, R.C. Walker, J. Wang, R.M. Wolf, X. Wu, D.M. York and P.A. Kollman (2015), AMBER 2015, University of California, San Francisco.

\bibitem{43}
Ch.W. Hopkins, S. Le Grand, R.C. Walker, A.E. Roitberg, Long-Time-Step Molecular Dynamics through Hydrogen Mass Repartitioning,
\newblock {\em J. Chem. Theory Comput.} 11 (20015) 1864-1874. https://doi.org/10.1021/ct5010406.

\bibitem{44}
E.F. Pettersen, T.D. Goddard, C.C. Huang, G.S. Couch, D.M. Greenblatt, E.C. Meng, T.E. Ferrin, UCSF Chimera--a visualization system for exploratory research and analysis,
\newblock {\em J. Comput. Chem.} 25 (2004) 1605-1612. https://doi.org/10.1002/jcc.20084.

\bibitem{45}
L.S.D. Caves, J.D. Evanseck, M. Karplus, Locally accessible conformations of proteins: Multiple molecular dynamics simulations of crambin,
\newblock {\em Protein Science} 7 (1998) 649-666. https://dx.doi.org/10.1002

\bibitem{46}
P.S. Shenkin, D.Q McDonald, Cluster analysis of molecular conformations,
\newblock {\em Journal of Computational Chemistry} 15 (1994) 899-916. https://doi.org/10.1002/jcc.540150811.

\bibitem{47}
M. Feig, J. Karanicolas, Ch. L. Brooks, III: MMTSB Tool Set, MMTSB NIH Research Resource,
\newblock {\em The Scripps Research Institute (2001).} 

\bibitem{48}
Gaussian 09, Revision A.1, M.J. Frisch, G.W. Trucks, H.B. Schlegel, G.E. Scuseria, M.A. Robb, J.R. Cheeseman, G. Scalmani, V. Barone, B. Mennucci, G.A. Petersson, H. Nakatsuji, M. Caricato, X. Li, H.P. Hratchian, A.F. Izmaylov, J. Bloino, G. Zheng, J.L. Sonnenberg, M. Hada, M. Ehara, K. Toyota, R. Fukuda, J. Hasegawa, M. Ishida, T. Nakajima, Y. Honda, O. Kitao, H. Nakai, T. Vreven, J. A. Montgomery, Jr., J.E. Peralta, F. Ogliaro, M. Bearpark, J.J. Heyd, E. Brothers, K.N. Kudin, V.N. Staroverov, R. Kobayashi, J. Normand, K. Raghavachari, A. Rendell, J.C. Burant, S.S. Iyengar, J. Tomasi, M. Cossi, N. Rega, J.M. Millam, M. Klene, J.E. Knox, J.B. Cross, V. Bakken, C. Adamo, J. Jaramillo, R. Gomperts, R.E. Stratmann, O. Yazyev, A.J. Austin, R. Cammi, C. Pomelli, J.W. Ochterski, R.L. Martin, K. Morokuma, V.G. Zakrzewski, G.A. Voth, P. Salvador, J.J. Dannenberg, S. Dapprich, A. D. Daniels, O. Farkas, J.B. Foresman, J.V. Ortiz, J. Cioslowski, and D.J. Fox, Gaussian, Inc., Wallingford CT, 2009.

\bibitem{49}
W. Humphrey, A. Dalke, K. Schulten, VMD: Visual Molecular Dynamics,
\newblock {\em J. Molec. Graphics} 14 (1996) 33-38. https://doi.org/10.1016/0263-7855(96)00018-5.

\bibitem{50}
D.G. Fedorov, T. Ishida, K. Kitaura, Multilayer Formulation of the Fragment Molecular Orbital Method (FMO),
\newblock {\em J. Phys. Chem. A} 109 (2005) 2638-2646. https://doi.org/10.1021/jp047186z.

\bibitem{51}
H. Autrup, F.A. Barile, S.C. Berry, B.J. Blaauboer, A. Boobis, H. Bolt, C.J. Borgert, W. Dekant, D. Dietrich, J.L. Domingo, G.B. Gori, H. Greim, J. Hengstler, S. Kacew, H. Marquardt, O. Pelkonen, K. Savolainen, P. Heslop-Harrison, N.P. Vermeulen, Human exposure to synthetic endocrine disrupting chemicals (S-EDCs) is generally negligible as compared to natural compounds with higher or comparable endocrine activity. How to evaluate the risk of the S-EDCs?,
\newblock {\em Toxicology Letters} 331 (2020) 259-264. https://doi.org/10.1016/j.toxlet.2020.04.008.

\bibitem{52}
J. Jin, P. Wu, X. Zhang, D. Li, W-L. Wong, Y-J. Lu, N. Sun, K. Zhang, Understanding the interaction of estrogenic ligands with estrogen receptors: a survey of the functional and binding kinetic studies,
\newblock {\em Journal of Environmental Science and Health, Part C} 38 (2020) 142-168. https://doi.org/10.1080/26896583.2020.1761204.

\bibitem{53}
A. Matsushima, A Novel Action of Endocrine-Disrupting Chemicals on Wildlife; DDT and Its Derivatives Have Remained in the Environment,
\newblock {\em Int. J. Mol. Sci.} 19 (2018) 1377-1390. https://doi.org/10.3390/ijms19051377.

\bibitem{54}
N.A. Whitman, Z-W. Lin, T.J. DiProspero, J.C. McIntosh, M.R. Lockett, Screening Estrogen Receptor Modulators in a Paper-Based Breast Cancer Model,
\newblock {\em Anal. Chem.} 90 (2018) 11981-11988. https://doi.org/10.1021/acs.analchem.8b02486.

\bibitem{55}
M. Liu, S. Zhang, S. Du, S. Pang, X. Liu, H. Zhang, A high throughput screening method for endocrine disrupting chemicals in tap water and milk samples based on estrogen receptor a and gold nanoparticles,
\newblock {\em Anal. Methods} 12 (2020) 200-204. https://doi.org/10.1039/c9ay02179e.

\bibitem{56}
G.M. Anstead, K.E. Carlson, J.A. Katzenellenbogen, The estradiol pharmacophore: Ligand structure-estrogen receptor binding affinity relationships and a model for the receptor binding site,
\newblock {\em Steroids} 62 (1997) 268-303. https://doi.org/10.1016/S0039-128X(96)00242-5.

\bibitem{57}
C.C. Valley, A. Cembran, J.D. Perlmutter, A.K. Lewis, N.P. Labello, J.Gao, J.N. Sachs, The Methionine-aromatic Motif Plays a Unique Role in Stabilizing Protein Structure,
\newblock {\em The Journal of Biological Chemistry} 287 (2012) 34979-34991. https://doi.org/10.1074/jbc.M112.374504.

\bibitem{58}
A. Jasanof, M.A. Weiss, Aromatic-Histidine Interactions in the Zinc Finger Motif Structural Inequivalence of Phenylalanine and Tyrosine in the Hydrophobic Core,
\newblock {\em Biochemistry} 32 (1993) 1423-1432. https://doi.org/10.1021/bi00057a005.

\bibitem{59}
E.Cauet, M. Rooman, R. Wintjens, J. Lievin, C. Biot, Histidine-Aromatic Interactions in Proteins and Protein-Ligand Complexes: Quantum Chemical Study of X-ray and Model Structures,
\newblock {\em J. Chem. Theory Comput.} 1 (2005) 472-483. https://doi.org/10.1021/ct049875k.

\bibitem{60}
L. Shimoni, J.P. Glusker, The Geometry of Intermolecular Interactions in Some Crystalline Fluorine-Containing Organic Compounds,
\newblock {\em Struct. Chem.} 5 (1994) 383-397. https://doi.org/10.1007/BF02252897.

\bibitem{61}
R.M. Blair, H. Fang, W.S. Branham, B.S. Hass, S.L. Dial, C.L. Moland, W. Tong, L. Shi, R. Perkins, D.M. Sheehan, The Estrogen Receptor Relative Binding Affinities of 188 Natural and Xenochemicals: Structural Diversity of Ligands,
\newblock {\em Toxicological Sciences} 54 (2000) 138-153. https://doi.org/10.1093/toxsci/54.1.138.

\bibitem{62}
M.K. Gilson, H-X. Zhou, Calculation of Protein-Ligand Binding Affinities,
\newblock {\em Annu. Rev. Biophys. Biomol. Struct.} 36 (2007) 21-42. https://doi.org/10.1146/annurev.biophys.36.040306.132550.

\bibitem{63}
M.A. Williams, Protein-Ligand Interactions: Fundamentals, in: M.A. Williams, T. Daviter (Eds.), Protein-Ligand Interactions: Methods and Applications Second Edition, Methods in Molecular Biology, vol. 1008, Springer Science+Business Media New York, 2013, pp. 3-34.
https://doi.org/10.1007/978-1-62703-398-5.

\end{thebibliography}

\end{document}